\documentclass[journal]{IEEEtran}
\IEEEoverridecommandlockouts
\usepackage{cite}
\usepackage{amsmath,amssymb,amsfonts}
\usepackage{algorithmic}
\usepackage{graphicx}
\usepackage{textcomp}
\usepackage{xcolor}
\usepackage[shortlabels]{enumitem}
\usepackage{hyperref}
\usepackage{xltabular}
\usepackage{gensymb}
\usepackage{multirow}

\def\BibTeX{{\rm B\kern-.05em{\sc i\kern-.025em b}\kern-.08em
    T\kern-.1667em\lower.7ex\hbox{E}\kern-.125emX}}
    
\DeclareMathOperator\erf{erfc}

\begin{document}

\title{Clipping noise cancellation receiver for the downlink of massive MIMO OFDM system}

\author{Marcin Wachowiak,~\IEEEmembership{Student Member,~IEEE,}
        and~Pawel~Kryszkiewicz,~\IEEEmembership{Senior~Member,~IEEE}% <-this % stops a space
\thanks{P. Kryszkiewicz is with the Institute of Radiocommunications, Poznan University of Technology, POLAND e-mail: pawel.kryszkiewicz@put.poznan.pl}% <-this % stops a space
\thanks{This research was funded by the Polish National Science Centre, project no. 2021/41/B/ST7/00136. For the purpose of Open Access, the author has applied a CC-BY public copyright license to any Author Accepted Manuscript (AAM) version arising from this submission.}% <-this % stops a space
\thanks{Manuscript received ...}}

\markboth{accepted to IEEE Transactions on Communications}%
{M. Wachowiak, P. Kryszkiewicz: Clipping noise cancellation receiver for the downlink of massive MIMO OFDM system}

\maketitle

\begin{abstract}
Massive multiple-input multiple-output (mMIMO) technology is considered a key enabler for the 5G and future wireless networks. In most wireless communication systems, mMIMO is employed together with orthogonal frequency-division multiplexing (OFDM) which exhibits a high peak-to-average-power ratio (PAPR). While passing the OFDM signal through one of the common RF front-ends of limited linearity, significant distortion of the transmitted signal can be expected. In mMIMO systems, this problem is still relevant as in some channels the distortion component is beamformed in the same directions as the desired signal. In this work, we propose a multi-antenna clipping noise cancellation (MCNC) algorithm for the downlink of the mMIMO OFDM system. Computer simulations show it can remove nonlinear distortion even under severe nonlinearity. Next, a simplified version of the algorithm is proposed. It was observed that for the direct visibility channels, its performance is only slightly degraded with respect to the MCNC algorithm.
\end{abstract}

\begin{IEEEkeywords}
orthogonal frequency-division multiplexing (OFDM), massive MIMO (mMIMO), front-end nonlinearity, clipping noise cancellation (CNC) 
\end{IEEEkeywords}

\section{Introduction}
Massive multiple-input multiple-output (mMIMO) systems are envisioned as the key enabler of the latest fifth generation of wireless networks and beyond. The high number of antennas combined with advanced signal processing allows an increase in the throughput to meet the growing demands. In \cite{Bjornson_2018_infinite_capacity} it was theoretically shown that the capacity of mMIMO systems is not upper-bounded and can be infinitely increased with the growing number of antennas. However, when considering practical implementation, hardware impairments, limiting the performance of the system, need to be taken into account. One of the crucial impairments to the transmit and receive signal chains is nonlinear amplification. Most terrestrial mMIMO systems employ the orthogonal frequency-division multiplexing (OFDM) technique due to its high bandwidth efficiency and low-complexity receiver structure. However, OFDM modulation is characterized by a high peak-to-average-power ratio (PAPR) \cite{Gharaibeh_distortion_book}, which combined with nonlinear amplification results in significant nonlinear distortion of the signal.

With the advent of massive MIMO communications, the problem of nonlinear distortion reappeared in a new context. The presence of nonlinearity in multiple antenna systems introduces an additional degree of complexity, which has to be carefully considered. Initial analyses \cite{bjornson_nonlinearity_as_white_noise} assumed that the distortion can be modeled as additive white noise uncorrelated between antennas. However, this work considered narrowband transmission on a single carrier. Later, the analysis in \cite{oob_clarified_two_tone_test} has proven that the distortion signals are in some scenarios correlated among antennas. The analysis was performed in a multiple antenna system with two subcarriers and a nonlinearity modeled as a third-order polynomial. A follow-up work \cite{on_antenna_array_oob}, which included the OFDM waveform, also found that some in-band and out-of-band emissions are always beamformed in the same directions as the desired signals, i.e., an increase in the number of transmitting antennas does not increase the signal to distortion power ratio (SDR). In \cite{mollen_spatial_char}, a detailed study of the radiation characteristic of the distortion signal was performed, addressing also OFDM signals. The authors derived a spatial cross-correlation matrix of nonlinear distortion components, which can be used to predict the expected signal-to-distortion levels, both in-band and out-of-band. In \cite{mollen_oob_radiation_large_arrays}, it was found, for signals with a high peak-to-average power ratio (PAPR), that with the growing number of users being served simultaneously, the distortion signal radiation characteristic becomes approximately omnidirectional. However, for direct visibility channels and a single user, SDR remains constant regardless of the number of antennas. This points to the conclusion that nonlinear distortion is still a major impairment even in mMIMO systems and measures must be taken to mitigate its effects on the system performance.

In single-input single-output (SISO) systems utilizing OFDM, several solutions to the nonlinear front-end problem have been proposed at the transmitter side \cite{papr_reduction_techniques}. One commonly employed technique is clipping and filtering (CAF) presented in \cite{clipping_and_filtering}. It allows for PAPR reduction without average power increase or bandwidth broadening. One critical issue of CAF is the presence of in-band distortion originating from the clipping. In the literature, two distinguished approaches toward distortion recovery and removal at the receiver can be found: time-domain (TD) and frequency-domain (FD). The TD approach is represented by decision-aided reconstruction (DAR) \cite{decision_aided_reconstruction} and the FD approach by clipping noise cancellation (CNC) \cite{original_cnc_algorithm}. In \cite{ochiai_cnc_and_dar_eval} it was shown that the CNC algorithm outperforms DAR, which was supported by the derivation of theoretical performance bounds.

% Similarly in mMIMO OFDM systems, the nonlinear distortion can be combated both at the transmitter side via precoding or at the receiver side via distortion cancellation methods.
% In \cite{z3ro_precoder} authors have proposed a novel precoding method for large arrays, which cancels the coherent combining of the third-order nonlinear distortion at the receiver. The precoder is also applicable for OFDM-modulated signals. In \cite{dab_for_mmw_miso} a distortion-aware iterative linear precoder for a MISO millimeter-wave channel has been proposed. It allows to effectively null the distortion in the desired direction for a high-SNR regime, single-user case and any transmit waveform.

So far, mMIMO OFDM receivers aware of nonlinear distortion have received limited attention in the literature. 
In \cite{performance_of_distortion_aware_linear_receivers_in_mimo} authors have derived and analyzed the performance of  a distortion-aware linear minimum mean squared error-based receiver for the uplink in an mMIMO OFDM system. The receiver offers some performance improvement, however, it is still far from reaching the performance of a system without nonlinear amplification.
In \cite{cs_based_nonlinearity_compensation} compressive sensing is used together with an orthogonal matching pursuit algorithm to compensate for the nonlinearity in the receiver at the base station. The method is evaluated for an mMIMO OFDM system with the Saleh model of a nonlinear amplifier. The results are compared against a neural network compensator, both at the receiver and transmitter.
In \cite{ndic_algorithm_for_mmimo} a joint channel equalization and iterative nonlinear distortion cancellation technique are discussed for the uplink in a Multi-User mMIMO system. The utilized algorithm is very similar to the CNC, however, it was analyzed for a single carrier transmission.
In \cite{pa_nonlinearity_cancellation} authors propose a power amplifier noise cancellation (PANC) algorithm for the uplink in a multi-user space division multiple access (SDMA) OFDM system. While its principle of operation is similar to the CNC algorithm, the considered scenario, i.e., multiple single antenna nonlinear transmitters delivering signal to a linear, multi-antenna receiver, is significantly different from the one considered in this paper. The performance of the algorithm is evaluated with joint channel estimation. Additionally, an upper bound bit error rate (BER) is derived subject to the considered system parameters.
In \cite{cnc_for_ofdm_oam_systems} the CNC algorithm is studied for an orbital angular momentum (OAM) multiplexing system with a uniform circular array both at the receiver and transmitter. The work considers a line-of-sight channel with OAM beamforming. A learning-based distortion recovery algorithm is presented. It resembles the CNC algorithm in its unfolded form with the introduction of additional learnable parameters which have to be optimized. It is important to mention that nonlinear distortions introduce some additional frequency diversity allowing for reception quality higher than in the linear OFDM case at the cost of increased computational complexity. A generalized approximate message passing algorithm is used for this purpose in \cite{belief_propagation_rx_ofdm} for a SISO OFDM system. In \cite{mimo_ofdm_gamp_rx} the scheme was applied to a singular value decomposition (SVD)-based MIMO OFDM system to combat digital-to-analog converter (DAC) nonlinearity distortion. The listed works mostly address the problem of nonlinear distortion in the uplink of an mMIMO OFDM system. Therefore, the precoding and combining of the signals from multiple antennas are not considered.

In this work, we focus on a single-user downlink transmission in a massive MIMO OFDM system. It corresponds to the worst-case scenario when SDR is the lowest due to the distortion being beamformed in the same direction as the desired signal \cite{oob_clarified_two_tone_test}. We propose a multi-antenna clipping noise cancellation algorithm (MCNC), which takes into consideration precoding and propagation in a multi-antenna system. Introduced reconstruction of the transmit chain in the MCNC algorithm is required for effective cancellation of the distortion in multi-antenna scenarios. Then a simplified receiver is derived for a specific precoding case. It requires fewer computations and control information and resembles the standard CNC algorithm used for SISO systems. The performance of the algorithms is evaluated for MRT precoding and a few channel models. The simulation results allow for a comparison of the algorithms in regard to a number of parameters.

The main contributions of this work are as follows: 
1) Justification of a complex-Gaussian distribution of OFDM symbol samples after precoding allowing the use of results for OFDM signal decomposition.
2) Evaluation of the influence of the channel type (LOS, two-path, IID Rayleigh), the number of antennas and the power amplifier (PA) input back off (IBO) on SDR under maximum ratio transmission (MRT) precoding.
3) A new MCNC algorithm is proposed for the removal of clipping noise in the receiver of the downlink mMIMO OFDM system, designed to effectively cancel the distortion from multiple transmit antennas. 
4) A simplified version of the MCNC algorithm is proposed performing close to the MCNC algorithm for channels with limited frequency selectivity. 
5) The scheme's performance is verified in various channels, i.e., line of sight (LOS), two-path and independent, identically distributed (IID) Rayleigh and system configurations. Additionally, the influence of channel coding, 3GPP 38.901 channel model\cite{3gpp_nr_channel_model}, and imperfect channel estimation have been considered. The convergence has been analyzed both in terms of the required signal quality and the number of iterations.

The remainder of this paper is organized as follows. Section \ref{sec:system_model} describes the mMIMO OFDM transmission system and the iterative receivers. Then the computational complexity of proposed algorithms is discussed in Sec. \ref{sec:computational_complexity}. The simulation results are presented in Sec. \ref{sec:simulation_results}. Finally, the concluding remarks are given in Sec. \ref{sec:conclusions}.

\section{System model}
\label{sec:system_model}
\begin{figure}[htb]
\centering
\includegraphics[width=3.6in]{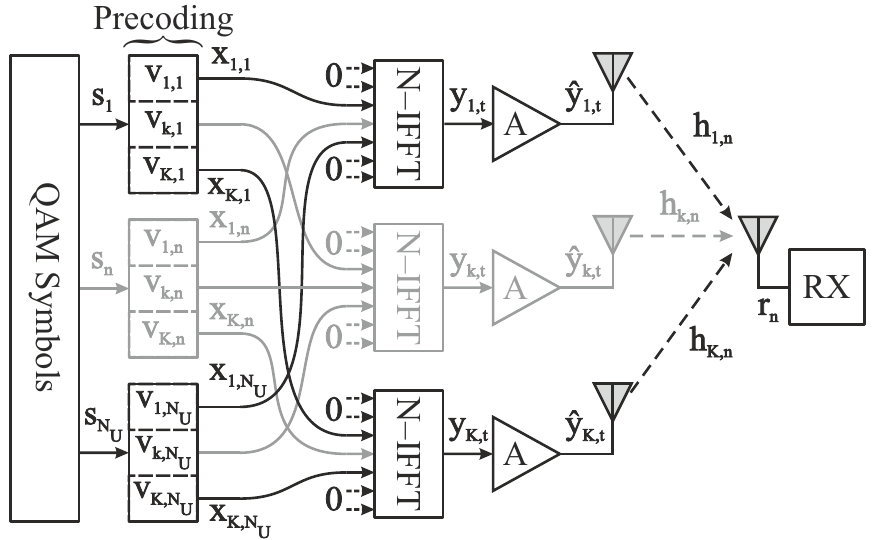}
\caption{System model.}
\label{fig:system_model}
\end{figure}

An mMIMO OFDM transmission system depicted in Fig. \ref{fig:system_model} is considered. There are $N_{\mathrm{U}}$ quadrature amplitude modulation (QAM) symbols $s_n$ ($n\in\{1,...,N_{\mathrm{U}}\}$ transmitted over adjacent subcarriers in a single OFDM symbol period. The symbols are chosen from set $\chi$. The symbols are precoded and transmitted by $K$ parallel transmitting signal chains, each consisting of an OFDM modulator with a maximum number of $N$ subcarriers, a nonlinear amplifier and an antenna element. Signals from different antennas combine at the single antenna receiver.

\subsection{Radio channel}
In order to utilize the OFDM modulator, it is assumed that the radio channel is constant for the frequency span of a single subcarrier, i.e., channel coherence bandwidth is not smaller than a single subcarrier bandwidth. For $n$-th subcarrier and $k$-th antenna, the channel response is a single complex coefficient expressed as $h_{k,n}$.

\subsection{Precoding}
Precoding is applied by multiplying the data symbol at $n$-th subcarrier $s_n$ by precoding coeffcient $v_{k,n}$ for $n$-th subcarrier and $k$-th antenna obtaining the precoded symbol $x_{k,n}$:
\begin{equation}
\label{eq:precoded_symbols}
    x_{k,n} = s_n v_{k,n}.
\end{equation}

It is assumed that the precoder is normalized to obtain a unitary summarized transmit power gain, irrespective of the number of utilized antennas for each subcarrier independently, i.e.,
\begin{equation}
\label{eq:precod_normalization}
    \sum_{k=1}^{K}\left| s_n v_{k,n}\right|^2=\left| s_n\right|^2\sum_{k=1}^{K}\left| v_{k,n}\right|^2=\left| s_n\right|^2.
\end{equation}
For a special case of MRT, which maximizes the received power, the precoding coefficients are calculated as\cite{mmimo_fundamentals_marzetta_2016}:
\begin{equation}
\label{eq:mrt_precoding}
        v_{k,n} = \frac{h_{k,n}^*} {\sqrt{\sum_{\tilde{k}=1}^{K}{\left| h_{\tilde{k},n}\right|^2}}},
\end{equation}
where $*$ denotes complex conjugate. 

\subsection{OFDM Modulation}
Precoded symbols are then subject to OFDM modulation \cite{haykin_digital_comms}, which is performed by inverse fast Fourier transform (IFFT) of size $N$. Only $N_{\mathrm{u}}$ subcarriers of indices $\mathcal{N}$ are modulated by data symbols $x_{k,n}$. The other $N-N_{\mathrm{u}}$ subcarriers are modulated with zeros. Typically, for a symmetric OFDM spectrum and an unused direct current (DC) subcarrier the subcarrier indices set equals $\mathcal{N}=\{-N_{\mathrm{u}}/2,...,-1, 1,...,N_{\mathrm{u}}/2\}$.
At the output of the IFFT, the $t$-th sample of OFDM signal for $k$-th antenna is calculated as:
\begin{equation}
    y_{k,t} = \frac{1}{\sqrt{N}} \sum_{n\in \mathcal{N}}x_{k,n}e^{j2\pi\frac{n}{N}t}, 
    \label{eq:IFFT}
\end{equation}
where $t \in \{-N_{\mathrm{CP}},....,N-1\}$, and $N_{\mathrm{CP}}$ is the number of samples of the cyclic prefix (CP).

\subsection{Nonlinear amplifier}
\label{sec_Nonlinear_amplifier}
The modulated signal undergoes the standard digital-to-analog conversion and upconversion to a chosen carrier frequency. These steps are omitted in our model as they are reversed at the receiver. Next, the signal is subject to nonlinear amplification by a nonlinear amplifier model identical for each transmitting signal chain:
\begin{equation}
    \label{eq:processing_by_nonlinearity}
    \hat{y}_{k,t} = \mathcal{A}(y_{k,t}),
\end{equation}
which in the case of the soft limiter \cite{Gharaibeh_distortion_book} can be described as:
\begin{equation}
\label{tx_dist_sig}
\hat{y}_{k,t} = \begin{cases} y_{k,t} & \mathrm{for} \left|y_{k,t}\right|^2 \leq P_{\mathrm{max}} \\
                  \sqrt{P_{\mathrm{max}}}e^{j \arg{(y_{k,t})}} & \mathrm{for} \left|y_{k,t}\right|^2 > P_{\mathrm{max}} \end{cases} ,
\end{equation}
where $P_{\mathrm{max}}$ is the maximum transmit power of a given PA and $\arg{(y_{k,t})}$ denotes phase of $y_{k,t}$. If the instantaneous signal power exceeds the $P_{\mathrm{max}}$, the signal is clipped, i.e., has constant amplitude while maintaining the input phase. While there is a number of different PA models, the soft limiter is proved to be the nonlinearity maximizing the SDR \cite{Raich_2005_clipper_optimal}. While in many contemporary systems digital predistortion is employed, the soft limiter can be treated as an optimal characteristic of the combined PA-predistorter model.       

It is a common practice to use IBO to determine PA operating point and respectively the $P_{\mathrm{max}}$. It is defined as a ratio of maximum PA power to the average power at the input of the amplifier, expressed in decibel scale: 
\begin{equation}
\label{eq_IBO}
        IBO\ [dB] = 10log_{10}\left( \frac{P_{\mathrm{max}}}{
     \mathbb{E} [ |y_{k,t}|^{2} ]
        } \right),
\end{equation}
where the expectation operator is denoted as $\mathbb{E}$.
Assuming that the average signal power is calculated based on each OFDM symbol sample over all antennas and using (\ref{eq:precod_normalization}) we get:
\begin{equation}
    \label{eq:output_power_lvls}
    \mathbb{E} [ |y_{k,t}|^{2} ] =  \frac{\bar{P_s}}{NK} \sum_{n\in \mathcal{N}}\sum_{k=1}^{K}|v_{k,n}|^2=\frac{\bar{P_s}N_{\mathrm{u}}}{KN},
\end{equation}
where $\bar{P_s}$ is the average power of a single symbol $s_n$. If the wireless channel is varying in time the expectation over $|v_{k,n}|^2$ should also be considered.
%It is calculated as an average of symbol powers in a constellation set of size M:
%\begin{equation}
%\bar{P_s} = \frac{1}{M}\sum_{m=1}^{M}{|q_m|^2},
%\end{equation}
%where the $q_m$ denotes a constellation symbol of index $m$. 
Because of averaging mean power over antennas in (\ref{eq:output_power_lvls}), all $K$ amplifiers work with the same clipping threshold $P_{\mathrm{max}}$.

The signal at the output of the amplifier can be decomposed based on the principle of homogenous linear mean square estimation \cite{papoulis2002random} as:
%\cite{Rowe_distorted_sig_decomposition}:
% \PK{W wiekszosci artykulow jest odwolanie na teorie Busgannga, ale wlasciwie ona jest troche szersza, a wymaga, zeby sygnal wejsciowy byl Gaussowski- jak dlatego wole to uzasadnienie}
\begin{equation}
\label{eq:bussgang_decomposition}
\hat{y}_{k,t} = \alpha_k y_{k,t} + \bar{d}_{k,t}
\end{equation}
where $\alpha_k$ is the correlation coefficient specific for $k$-th antenna, $\bar{d}_{k,t}$ is the distortion signal uncorrelated with the desired signal $y_{k,t}$. The coefficient $\alpha_k$ is defined as follows:
\begin{equation}
\label{eq:lambda_estimation}
    \alpha_k = \frac{\mathbb{E}\left[ \hat{y}_{k,t} y^*_{k,t} \right]}{\mathbb{E}\left[ y_{k,t} y^*_{k,t}  \right]}.
\end{equation}

The value $\alpha_k$ can be derived analytically assuming the complex-Gaussian distribution of $y_{k,t}$ \cite{Rowe_distorted_sig_decomposition}. While an exact signal envelope distribution for QAM-modulated OFDM is of a discrete nature \cite{Yoo_2011_dist_OFDM}, it converges fast with the number of subcarriers to its limit, i.e., a complex-Gaussian distribution. This comes from 
the utilization of the central limit theorem as $N_{\mathrm{U}}\gg 0$ independently modulated subcarriers are used. In \cite{Wei_2010_dist_OFDM} it has been shown that the limit distribution is obtained not only for independent and identically distributed symbols. It is valid as well for coded systems, allowing the modulating symbols to be dependent but uncorrelated. Additionally, power variation among subcarriers, e.g., as a result of water filling, still allows the complex-Gaussian distribution to be used. These derivations allow the complex-Gaussian distribution to be assumed for the mMIMO OFDM signal. 
First, while various precoders $v_{k,n}$ can be used, e.g., MRT or zero-forcing (ZF) \cite{mmimo_fundamentals_marzetta_2016}, these typically depend on the wireless channel properties, not the modulating symbols resulting in $\forall_{n\in \mathcal{N}} \mathbb{E}[ s_n v_{k,n} ]=\mathbb{E}[ s_n] \mathbb{E}[v_{k,n} ]$. As such, using a common assumption that QAM symbols are uncorrelated of zero mean, i.e., $\forall_{n\neq m} \mathbb{E}[ s_n s_{m}^{*} ]=\mathbb{E}[ s_n] \mathbb{E}[s_{m}^{*} ]$ and $\mathbb{E}[s_n]=0$, it can be shown that
\begin{equation}
    \forall_{n\neq m} \mathbb{E}[ x_{k,n} x_{k,m}^{*} ]=\mathbb{E}[ s_n] \mathbb{E}[ s_m]^{*} \mathbb{E}[v_{k,n}v_{k,m}^{*} ]=0.
\end{equation}
Therefore, the symbols $x_{k,n}$ are uncorrelated as required by \cite{Wei_2010_dist_OFDM}. 
The second issue is the power variation among subcarriers. It can happen as a result of some sort of water filling, resulting in $\exists_{m\neq n} \mathbb{E}[|s_n|^2]\neq \mathbb{E}[|s_m|^2]$.  However, it is possible that power amplification by coefficient $v_{k,n}$ can vary among subcarriers, e.g., in the case of MRT precoder as a result of frequency selective fading. Still, \cite{Wei_2010_dist_OFDM} shows the complex-Gaussian assumption can be used in these cases.  

As such $\alpha_k$ can be calculated as in \cite{Rowe_distorted_sig_decomposition} considering that power can be unequally distributed among antennas, e.g., as a result of some antenna array elements being pointed in a different direction than the served user, resulting in the increased power of other matrix elements for an MRT precoder described by (\ref{eq:mrt_precoding}). In the case of a common maximal transmit power $P_{\mathrm{max}}$ for all utilized front-ends, mean transmit (TX) power per antenna can be different resulting in varying per-antenna IBO, i.e.,
\begin{equation}
      IBO_k\ [dB] = 10log_{10}\left( \frac{P_{\mathrm{max}}}{
     \frac{\bar{P_s}}{N} \sum_{n\in \mathcal{N}}|v_{k,n}|^2
        } \right).
\end{equation}
% \MW{In simulator as:
% \begin{equation}
%       IBO_k\ [dB] = 10log_{10}\left( \frac{10^{\frac{IBO}{10}}N_{\mathrm{u}}}{K
%       \sum_{n\in \mathcal{N}}|v_{k,n}|^2
%         } \right).
% \end{equation}}
The $\alpha_k$ coefficient  can be calculated as\cite{Rowe_distorted_sig_decomposition}:
\begin{equation}
\label{eq:analytical_alpha}
    \alpha_k = 1 - e^{-\gamma_k^2} + \frac{\sqrt{\pi\gamma_k}}{2}\erf{\left(\gamma_k\right)},
\end{equation}
where $\gamma_k = 10^{\frac{IBO_k}{20}}$ and $\erf(\cdot)$ denotes the error function. Observe that in many architectures and for many channel types the coefficient $\alpha_k$ will be invariant with respect to the antenna index as a result of equal power per antenna.

\subsection{Signal reception}
\label{sec_RX}
The signal transmitted in time domain $\hat{y}_{k,t}$ from $k$-th antenna is convolved with its respective wideband channel impulse response. After passing through the channel the $K$ signals are summed at the receiving antenna. After the removal of CP, the fast Fourier transform (FFT) is applied which allows to express the signal received at $n$-th subcarrier as:
\begin{equation}
\label{rx_sig_freq_domain}
    r_n=\sum^{K}_{k=1} \mathcal{F}_{[n,t=0,...,N-1]}\{\hat{y}_{k,t}\}h_{k,n} + w_n,
\end{equation}
where $w_n$ is the white noise sample at $n$-th subcarrier in the receiver and $\mathcal{F}_{[n,t=0,...,N-1]}\{ \cdot\}$ denotes discrete Fourier transform (DFT) over time instants $t=0,...,N-1$ at $n$-th subcarrier.

Based on (\ref{eq:bussgang_decomposition}) and (\ref{eq:IFFT}) the received signal can be expanded to:
\begin{equation}
\label{eq_RX_signal}
    r_n= \sum^K_{k=1}\alpha_k h_{k,n} x_{k,n} +\sum^K_{k=1} h_{k,n}d_{k,n}+w_n, 
\end{equation}
where 
\begin{equation}
    d_{k,n}=\mathcal{F}_{[n,t=0,...,N-1]}\{\bar{d}_{k,t}\}.
\end{equation}
Observe that in general $d_{k,n}$ for a single subcarrier depends on the transmitted symbols $s_n$ and precoding coefficients $v_{k,n}$ for all the utilized subcarriers $n\in \mathcal{N}$. This can be easily shown by treating the OFDM signal as a set of subcarriers undergoing intermodulation on a polynomial-modeled PA \cite{Bohara_2007_OFDM_IMD}. 

Taking into account the precoding coefficients definition in (\ref{eq:precoded_symbols}) it is obtained that
\begin{equation}
\label{eq:rx_sig}
    r_n=\sum^K_{k=1}\alpha_kh_{k,n}v_{k,n}s_{n} +\sum^K_{k=1} h_{k,n}d_{k,n}+w_n.
\end{equation}

The signal-to-noise ratio (SNR) is defined considering only the data-carrying subcarriers with the wanted signal attenuated by coefficients $\alpha_k$ giving 
% \begin{equation}
% \label{eq:snr}
%     SNR =\frac{
%     %\overbrace{
%      \bar{P_s} \sum^K_{k=1} \alpha_k^2
%     \left(\frac{1}{N_{\mathrm{u}}}\sum_{n\in \mathcal{N}}\left|h_{k,n} v_{k,n}\right|^2\right)
%     %}^{Desired\ signal}
%     }{\mathbb{E}\left[\left|w_n\right|^2\right]}.
% \end{equation}

\begin{equation}
\label{eq:snr}
    SNR =\frac{
    %\overbrace{
     \bar{P_s} \frac{1}{N_{\mathrm{u}}}
    \sum_{n\in \mathcal{N}}\left|\sum^K_{k=1}\alpha_k h_{k,n} v_{k,n}\right|^2
    %}^{Desired\ signal}
    }{\mathbb{E}\left[\left|w_n\right|^2\right]}.
\end{equation}

Based on the SNR definition the Eb/N0 can be calculated as:
\begin{equation}
\label{eq:ebn0}
\frac{Eb}{N0} =  \frac{SNR}{\log_2M},
\end{equation}
where M is the size of the constellation, i.e., the number of elements in set $\chi$.

Similarly, the SDR is defined considering only the data-carrying subcarriers:
% \begin{equation}
%     \label{eq:sdr}
%     SDR = \frac{
%   \bar{P_s} \sum^K_{k=1} \alpha_k^2
%     \sum_{n\in \mathcal{N}}\left|h_{k,n} v_{k,n}\right|^2
%     }{\sum^K_{k=1}\sum_{n\in \mathcal{N}} \left|h_{k,n}d_{k,n}\right|^2}.
% \end{equation}

\begin{equation}
    \label{eq:sdr}
    SDR = \frac{
    \bar{P_s} \sum_{n\in \mathcal{N}}\left|\sum^K_{k=1} \alpha_k h_{k,n} v_{k,n}\right|^2}{\sum_{n\in \mathcal{N}} \left|\sum^K_{k=1} h_{k,n}d_{k,n}\right|^2}.
\end{equation}

\subsection{Simple reception}
In a simple receiver, first an equalization is performed, e.g., ZF, dividing received symbol $r_n$ by $\sum^K_{k=1}\alpha_k h_{k,n}v_{k,n}$, effectively removing the effects of channel, precoding and nonlinearity on wanted signal, i.e.,
\begin{equation}
\label{eq:simple_detection}
    g_n=s_{n} +\frac{\sum^K_{k=1} h_{k,n}d_{k,n}}{\sum^K_{k=1}\alpha_k h_{k,n}v_{k,n}}+\frac{w_n}{\sum^K_{k=1}\alpha_k h_{k,n}v_{k,n}}.
\end{equation}
However, this results in scaling of distortion and white noise terms. 
The detection is performed by finding the closest symbol from the constellation set:
\begin{equation}
        \label{eq:hard_symbol_detection}
        \tilde{s}_n=\min_{s\in \chi}\left|s-g_n\right|^2.
    \end{equation}

\subsection{Multiple antenna clipping noise cancelation receiver (MCNC)}
\begin{figure}[htb]
\centering
\includegraphics[width=3.6in]{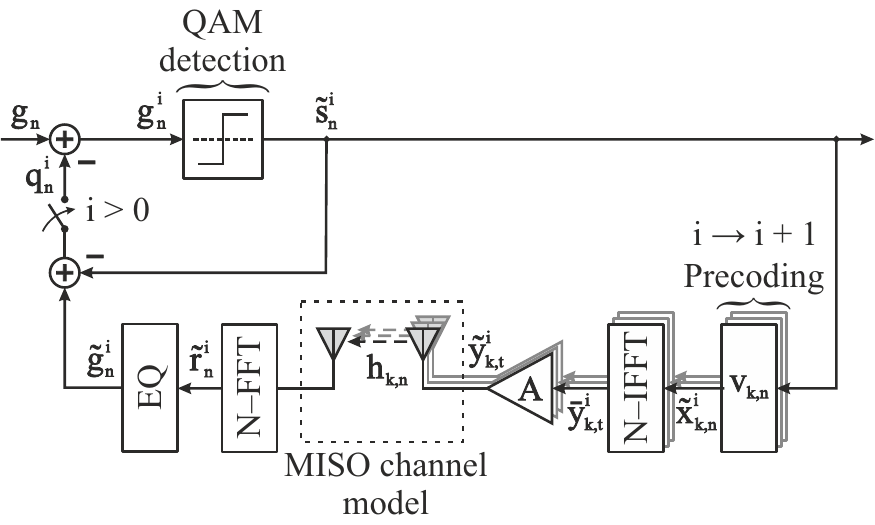}
\caption{Multiple antenna clipping noise cancellation algorithm flowchart.}
\label{fig:mcnc_flowchart}
\end{figure}

While the nonlinear distortion is often treated as white noise \cite{bjornson_nonlinearity_as_white_noise}, for the soft limiter it depends on the transmitted signal as shown in (\ref{tx_dist_sig}). Therefore, a decision-aided receiver is proposed that iteratively reproduces the received and nonlinearly distorted signal, improving detection quality. While the general idea is well known for SISO OFDM systems\cite{original_cnc_algorithm}, the mMIMO precoding and utilization of multiple antennas required it to be redesigned. The Multiple antenna CNC receiver is shown in Fig. \ref{fig:mcnc_flowchart}.

%Prior to the M(?)-CNC algorithm the received signal from (\ref{eq:rx_sig}) undergoes equalization to compensate for the attenuation and phase rotation introduced by the channel, that has not been taken into account with precoding

%\begin{equation}
%\label{eq:rx_sig_after_equalization}
%    \hat{r}_n=\frac{\sum^K_{k=1}\alpha_kh_{k,n}v_{k,n}s_{n}}{\sum^K_{k=1}h_{k,n}v_{k,n}} +\frac{\sum^K_{k=1} h_{k,n}d_{k,n}}{\sum^K_{k=1} h_{k,n}v_{k,n}}+\frac{w_n}{\sum^K_{k=1} h_{k,n}v_{k,n}}.
%\end{equation}

%\MW{ z uproszczeniem $\alpha$
 %   \begin{equation}
  %      \hat{r}_n= \alpha s_{n}+\frac{\sum^K_{k=1} h_{k,n}d_{k,n}}{\sum^K_{k=1} h_{k,n}v_{k,n}}+\frac{w_n}{\sum^K_{k=1} h_{k,n}v_{k,n}}.
   % \end{equation}
%}
%In some cases along with the desired signal power the distortion and noise components might also be amplified.

It consists of the following steps:
\begin{enumerate}[(a)]
    \item \label{cnc_hard_detection} 
    Hard symbol detection is performed for $n$-th subcarrier based on the received and equalized signal  $g_n^i$ with removed $i$-th nonlinearity distortion estimate where $i$ denotes the iteration number. For the $i=0$ the input is the original received signal $g_n$ as defined in (\ref{eq:simple_detection}). In the next iterations, the nonlinear distortion will be estimated and subtracted from $g_n$ constituting $g_n^i$.  
    
    The symbol detection is carried out by finding the closest, from a Euclidean distance perspective, symbol from the chosen QAM constellation set $\chi$: 
    \begin{equation}
        \label{eq:hard_symbol_detection_CNC}
        \tilde{s}^{i}_n = \arg \min_{s\in \chi}\left|s-g^i_n\right|^2.
    \end{equation}

    \item Obtained symbol estimate $\tilde{s}^{i}_n$ is used to regenerate the received signal using the whole link model including multiple antenna transmitters with nonlinear amplifiers, channel model and receiver with equalization. To achieve this the precoding and channel coefficients need to be known at the receiver. 
    
    First, the symbol estimate is precoded as in (\ref{eq:precoded_symbols}) using the same precoding coefficients:
    \begin{equation}
        \label{eq:precoded_symbol_estimate}
        \tilde{x}^i_{k,n} = \tilde{s}_n^i v_{k,n}.
    \end{equation}
    
    Then, the precoded symbol estimate is OFDM modulated as in (\ref{eq:IFFT}), using the same subcarrier mapping giving:
    \begin{equation}
        \label{eq:IFFT_estimate}
        \bar{y}^i_{k,t} = \frac{1}{\sqrt{N}} \sum_{n\in \mathcal{N}}\tilde{x}^i_{k,n}e^{j2\pi\frac{n}{N}t}. 
    \end{equation}
    
    Next, the signal is processed by the nonlinearity model as in (\ref{eq:processing_by_nonlinearity}) resulting in $\tilde{y}^i_{k,t}=\mathcal{A}(\bar{y}^i_{k,t})$.
    Signals obtained from each antenna are then passed through a multiple-input single-output (MISO) channel model similarly as in (\ref{rx_sig_freq_domain}), except for white noise addition, obtaining 
\begin{equation}
\label{rx_sig_freq_domain_iter}
    \tilde{r}_n^i=\sum^{K}_{k=1} \mathcal{F}_{[n,t=0,...,N-1]}\{\tilde{y}_{k,t}^i\}h_{k,n},
\end{equation}
    which is the regenerated received signal after the channel. If all the symbols $\tilde{s}_n^i$ are correct both the wanted signal and nonlinear distortion will be perfectly reconstructed. While this is not probable under severe nonlinearity or noise if most of the symbols $\tilde{s}_n^i$ are detected correctly the majority of nonlinear distortion should be reconstructed as well\cite{original_cnc_algorithm}.   
    
    The regenerated signal can be decomposed into desired and distortion components based on (\ref{eq:bussgang_decomposition}) as:
    \begin{equation}
        \tilde{r}^i_n= \sum^K_{k=1}\alpha_k h_{k,n} v_{k,n}\tilde{s}^i_{n} +\sum^K_{k=1} h_{k,n}\tilde{d}^i_{k,n}, 
    \end{equation}
    where $\tilde{d}^i_{k,n}$ denotes the reconstructed distortion signal received from $k$-th antenna on $n$-th subcarrier in $i$-th iteration.
    The regenerated signal undergoes equalization by dividing the signal by $\sum^K_{k=1} \alpha_k h_{k,n}v_{k,n}$ giving
    \begin{align}
        \label{eq_g_n_i_decomp}
        \nonumber \tilde{g}^i_n&=\frac{\tilde{r}^i_n}{\sum^K_{k=1} \alpha_k h_{k,n}v_{k,n}}
        \\&
        =\tilde{s}^i_{n} +\frac{\sum^K_{k=1} h_{k,n}\tilde{d}^i_{k,n}}{\sum^K_{k=1} \alpha_k h_{k,n}v_{k,n}}.
    \end{align}
The last component in  (\ref{eq_g_n_i_decomp}) is nonlinear distortion influencing $n$-th subcarrier if symbols $\tilde{s}^i_{n}$ were transmitted. While both $\tilde{g}^i_n$ and $\tilde{s}^i_{n}$ are known at this stage this signal can be calculated as 
    
    \begin{equation}
        \label{eq:regenerated_distortion_sig}
        q_n^i = \tilde{g}_n^i - \tilde{s}_n^i.
    \end{equation}
   \item The estimated distortion component is subtracted from the originally received signal
   \begin{equation}
    \label{eq:refined_rx_signal_simple}
       g_n^{i+1} = g_n - q_n^i
   \end{equation}
   constructing potentially improved received signal that can be used for detection in the next iteration.
   The algorithm returns to step \ref{cnc_hard_detection}  and repeats until a certain number of iterations has been reached or satisfactory quality of the received data has been achieved.
   
Using (\ref{eq:simple_detection}) and (\ref{eq_g_n_i_decomp}) the components of $g_n^{i+1}$ can be shown as:
      \begin{equation}
   \label{eq:refined_rx_signal_expanded}
      g_n^{i+1}= s_{n}+\frac{\sum^K_{k=1} h_{k,n}\left(d_{k,n}-\tilde{d}^i_{k,n}\right)+w_n}{\sum^K_{k=1} \alpha_k h_{k,n}v_{k,n}}.
   \end{equation}
   If the $\tilde{s}_n^i$ estimates are good enough, the estimated nonlinear distortion term $\tilde{d}^i_{k,n}$ should reduce the received distortion term $d_{k,n}$ improving the reception performance. 
   
   %\MW{ Z uproszczeniem $\alpha$
   %\begin{equation}
   %\label{eq:refined_rx_signal_expanded}
    %  \hat{r}_n^{i=i+1}= \alpha s_{n}+\frac{\sum^K_{k=1} h_{k,n}d_{k,n}-\sum^K_{k=1} h_{k,n}\tilde{d}^i_{k,n}}{\sum^K_{k=1} h_{k,n}v_{k,n}}+\frac{w_n}{\sum^K_{k=1} h_{k,n}v_{k,n}}.
   %\end{equation}}
\end{enumerate}
One of the disadvantages of the above algorithm is the requirement to know the channel coefficients and the precoding vectors used at the transmitter. This can be difficult in time division duplex (TDD)-based massive MIMO system in which channel reciprocity property is used \cite{mmimo_fundamentals_marzetta_2016}. In such case transmission of channel coefficients $h_{k,n}$ together with the utilized precoding coefficients $v_{k,n}$ will require a significant capacity of the control channel, especially for a high number of antennas and a frequency selective channel. Moreover, these coefficients have to be timely delivered in order not to delay the MCNC operation. 
%The algorithm starts once all time-domain samples of the OFDM frame have been received. It does not utilize any delay in the reception or the processing loop. The only requirement is the availability of up-to-date PA parameters, precoding and channel coefficients, which have to be acquired before the algorithm is executed. 

\subsection{CNC}
Considering the above-mentioned drawbacks of MCNC it is reasonable to propose a simplification resulting in lower computational complexity and a lower amount of control information required at the receiver.  

%\subsubsection{Flat channel - analog precoding}
An example that we start with is a precoder being fixed for all subcarriers of a given antenna. Moreover, we assume the precoder amplitude for each antenna is equal, that considering (\ref{eq:precod_normalization}), results in $|v_{k,n}|=\frac{1}{\sqrt{K}}$. 
Therefore, the precoding coefficient equals
\begin{equation}
\label{eq:fixed_precoder_all_sc}
    v_{k,n}=\frac{1}{\sqrt{K}}e^{j\varphi_k},
\end{equation}
where $\varphi_k$ is precoder phase shift specific for the $k$-th antenna. 
This allows to simplify (\ref{eq:IFFT}) as follows:
\begin{equation}
    y_{k,t} = \frac{1}{\sqrt{K}}e^{j\varphi_k}
    \underbrace{\frac{1}{\sqrt{N}} \sum_{n\in \mathcal{N}}s_{n}e^{j2\pi\frac{n}{N}t}}_{\ddot{y}_t}. 
\end{equation}
By combining (\ref{eq_IBO}) and (\ref{eq:output_power_lvls}) the clipping power of the considered PA can be defined as
\begin{equation}
    P_{\mathrm{max}} = 10^{\frac{\mathrm{IBO}}{10}}\frac{\bar{P_s}N_{\mathrm{u}}}{KN}.
\end{equation}
The precoded signal after the nonlinearity (\ref{tx_dist_sig}) can be rewritten as:
\begin{equation}
\label{tx_dist_sig_s_CNC}
\hat{y}_{k,t} = 
\begin{cases}
\frac{1}{\sqrt{K}}e^{j\varphi_k} \ddot{y}_t & 
\mathrm{for} \left|\frac{1}{\sqrt{K}} \ddot{y}_t\right|^2 \leq P_{\mathrm{max}} \\
\sqrt{P_{\mathrm{max}}}
e^{j\varphi_k +j\arg{(\ddot{y}_t)}} & 
\mathrm{for} \left|\frac{1}{\sqrt{K}} \ddot{y}_t\right|^2 > P_{\mathrm{max}} 
\end{cases}.
\end{equation}
This can be reformulated by taking the precoding coefficient as a common multiplier giving:
\begin{equation}
%\label{tx_dist_sig_s_CNC}
\hat{y}_{k,t} = \frac{1}{\sqrt{K}}e^{j\varphi_k}
\begin{cases}
 \ddot{y}_t & 
\mathrm{for} \left| \ddot{y}_t\right|^2 \leq \ddot{P}_{\mathrm{max}} \\
\sqrt{\ddot{P}_{\mathrm{max}}}
e^{j\arg{(\ddot{y}_t)}} & 
\mathrm{for} \left|\ddot{y}_t\right|^2 > \ddot{P}_{\mathrm{max}} 
\end{cases},
\end{equation}
where $\ddot{P}_{\mathrm{max}}=10^{\frac{\mathrm{IBO}}{10}}\frac{\bar{P_s}N_{\mathrm{u}}}{N}$ is clipping power for $K=1$ antenna. The last formula shows that for the considered precoder the signal for all antennas can be obtained by passing OFDM symbols obtained without precoding through a single PA (of the same IBO parameter) and scaling the PA output by the precoding coefficient (\ref{eq:fixed_precoder_all_sc}) before transmission through the $k$-th antenna. In other words, the considered nonlinear amplifier is equivalent to a linear operator for the precoding coefficient.
Similarly as in (\ref{eq:bussgang_decomposition}) the input to $k$-th antenna at time instance $t$ can be decomposed as:
\begin{equation}
    \hat{y}_{k,t} = \frac{1}{\sqrt{K}}e^{j\varphi_k} \left( \alpha \ddot{y}_{t}+\ddot{\bar{d}}_{t}\right),
\end{equation}
where $\ddot{\bar{d}}_{t}$ is distortion sample at time instance $t$ after passing $\ddot{y}_t$ through the nonlinear PA. Observe the distortion is independent of the precoding and the antenna index. Most interestingly, $\alpha$ is equal for each antenna and is the same as if precoding is considered before IFFT as in the MCNC algorithm. This is the result of $\alpha$ being dependent only on the IBO value as shown in (\ref{eq:analytical_alpha}).  
Following the same reasoning as in Sec. \ref{sec_RX} we obtain the received signal at $n$-th subcarrier:
\begin{equation}
\label{eq:flat_channel_rx_sig}
    r_n=\alpha s_n \frac{1}{\sqrt{K}}\sum^K_{k=1} e^{j\varphi_k} h_{k,n} +\ddot{d}_{n} \frac{1}{\sqrt{K}}\sum^K_{k=1} e^{j\varphi_k} h_{k,n}+w_n,
\end{equation}
where 
\begin{equation}
    \ddot{d}_{n}=\mathcal{F}_{[n,t=0,...,N-1]}\left\{\ddot{\bar{d}}_{t}\right\}.
\end{equation}
Initial ZF equalization results in 
\begin{equation}
\label{eq:equalization_CNC}
    g_n=s_{n} +\frac{\ddot{d}_{n}}{\alpha}
    +\frac{w_n}{\frac{\alpha}{\sqrt{K}}\sum^K_{k=1} e^{j\varphi_k} h_{k,n}}.
\end{equation}
While the aim of the proposed reception method is to reconstruct the clipping noise (distortion) this becomes relatively simple in this case. The values of $\ddot{d}_{n}$ depend only on the transmitted symbols $s_n$ and the PA IBO value. There is no need to know the channel coefficients and precoders nor to reconstruct all $K$ transmission chains.

\begin{figure}[htb]
\centering
\includegraphics[width=3.6in]{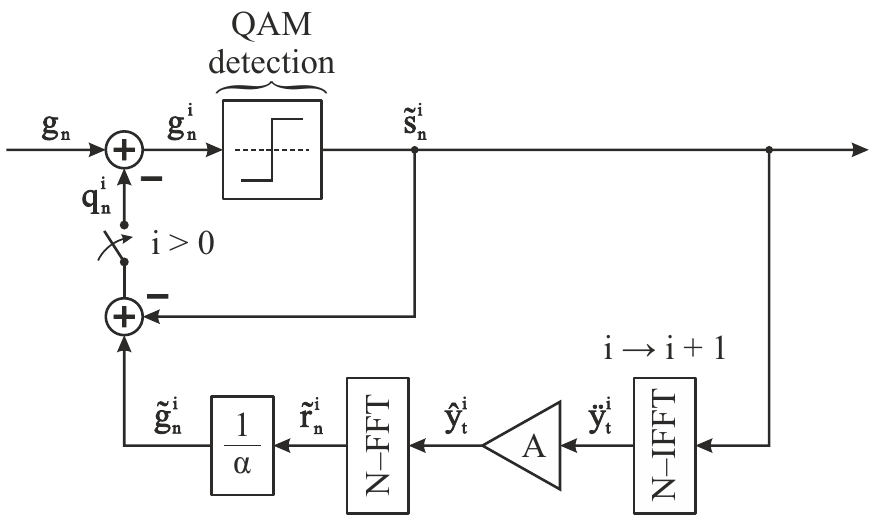}
\caption{Clipping Noise Cancellation algorithm flowchart.}
\label{fig:standard_cnc_flowchart}
\end{figure}

This allows us to propose a CNC algorithm for simplified reconstruction of clipping noise
in an mMIMO system, that has a similar structure as the algorithm described in \cite{ochiai_cnc_and_dar_eval} for single antenna systems. The signal processing flowchart of the CNC algorithm is shown in Fig. \ref{fig:standard_cnc_flowchart}. Its iterative structure is similar to the one used by MCNC except for using channel and precoding coefficients and multiple transmit antennas. 
The proposed CNC algorithm is based on a very specific precoding case. However, as will be supported by simulation results, it can be used as well for more complex precoding providing only slight performance degradation in comparison to the MCNC algorithm.  
% \PK{Tu troche sie zastanawiam bo na schemacie sa zmienne ktore beda liczone w inny sposob. Np. $\bar{y}_{k,t}$ nie ma sensu, bo to byla zmienna zalezna od indeksu anteny, a teraz nie ma tej zaleznsci. Dodatkowo wczesniej to y zalezalo od prekodowania, a w tym przypadku tylko od symboli $s_n$. Ale to wlasciwie trzebaby opisac caly algorytm CNC ponownie. Moze na cos wpadniemy.}

\subsection{Multi-user considerations}
\label{sec:multiuser}
The single-user scenario constitutes a worst-case from the nonlinear distortion power perspective as mentioned before \cite{mollen_spatial_char}.
However, an extension to a multi-user case should be discussed. This will require symbols of all users to be detected and used for nonlinear distortion reconstruction in each receiver. Even if all channel coefficients and precoding vectors could be obtained via some control channel, estimating other users' symbols would be challenging. Typically, simultaneously scheduled users have channels close to orthogonal. This results in a significantly attenuated wanted signal of the other simultaneously scheduled users at the considered user equipment. The SNR of the other users' signals will be much lower preventing successful detection. Additionally, the control and computational overhead  will be significant.

The other possibility is to use CNC/MCNC algorithms as described above. In this case, the signals of other users and part of the nonlinearity distortion will be treated as interference decreasing reception quality similarly to white noise. This will be one of the scenarios addressed in Sec. \ref{sec:simulation_results}.
%\PK{Dokonczyc}
%However, some use cases can be identified when the standard CNC and MCNC algorithms offer improvement. When the users' channels are significantly different in terms of attenuation, due to resource allocation and precoding the power of each user signal will also be different. This may correspond to the case when one user is close to the base station and the other is located far away from it. With the nonlinear distortion being a function of the power of the signal, the user with the dominating signal power will contribute most to the generated nonlinear distortion. Assuming that the other users' signal powers are small and negligible the CNC/MCNC algorithm can be effectively used for the user with dominating power at the transmitter.

\section{Computational complexity}
\label{sec:computational_complexity}
In this section, the computational complexity of a standard OFDM receiver, CNC and MCNC algorithms is analyzed in terms of real multiplications/divisions and additions/subtractions. It depends on the IFFT size $N$, the number of modulated subcarriers $N_\mathrm{U}$, the number of constellation points $M$, and the number of iterations of CNC/MCNC algorithm $I$. The FFT and IFFT is performed by radix-2 algorithm and requires $(N/2)\log_2{N}$ complex multiplications and $N\log_2{N}$ complex additions \cite{proakis_manolakis_dsp_book}. Each complex multiplication can be split into 3 real multiplications and 5 additions as shown in \cite{fast_algorithms_for_dsp}. With these simplifications, the FFT/IFFT operation cost is $3\left((N/2)\log_2{N}\right)$ real multiplications and $5\left((N/2)\log_2{N}\right)+ 2N\log_2{N}$ real additions. 

A single QAM symbol detection based on Euclidean distance (\ref{eq:hard_symbol_detection}) and by separating I/Q component requires $2\sqrt{M}$ comparisons, $2(2\sqrt{M})$ real multiplications and $2(3\sqrt{M})$ real additions, where $M$ is the constellation size. The OFDM symbol detection requires then $2N_\mathrm{U}\sqrt{M}$ comparisons, $4N_\mathrm{U}\sqrt{M}$ real multiplications and $6N_\mathrm{U}\sqrt{M}$ real additions. The precoding for a single front-end in a single-user case requires $\mathrm{N_U}$ complex multiplications, which translates to $3N_\mathrm{U}$ real multiplications and $5N_\mathrm{U}$ additions. A similar number of operations is required by the equalization and SISO channel propagation. Division by $\alpha$ coefficient requires two real divisions for each complex sample in $N_\mathrm{U}$ long vector.
 
Processing by a single nonlinear front-end requires $N$ comparisons, $2N$  multiplications and $N$ additions. When the sample power exceeds the $P_{\mathrm{max}}$ threshold it is multiplied by the square root of saturation power divided by the sample power. The CORDIC algorithm is employed to calculate the square root, which according to \cite{mathworks_cordic} requires 1 table lookup, 2 shifts and 3 real additions per iteration for a fixed point approximation. The number of iterations depends on the desired precision of the result, with each iteration corresponding to a single bit. Assuming the use of single precision floating arithmetic the number of iterations required by CORDIC is set to 23 \cite{cordic_accuracy}, resulting in 23 table lookups 46 shifts, and 69 real additions. This adds $2N$ real multiplications, $N$ divisions and $69N$ additions to the complexity of the operation. Table \ref{table:num_operations_dsp} presents a summarized number of operations for each signal processing step. 
%The IFFT size, which corresponds to the total number of subcarriers is denoted by $N$, among which the $N_{\mathrm{U}}$ represents the carriers utilized for data transmission.
\begin{table}[htb]
\centering
\caption{Number of operations of selected signal processing steps.}
\label{table:num_operations_dsp}
\bgroup
\def\arraystretch{1.2}
\begin{tabular}{ m{.23\columnwidth}  >{\centering\arraybackslash}  m{.31\columnwidth}  >{\centering\arraybackslash} m{.31\columnwidth}} 
 \hline
 \multirow{2}{*}{\begin{minipage}{.23\columnwidth}Signal processing step\end{minipage}} & \multicolumn{2}{c}{Operation count}  \\  \cline{2-3}
                        & Additions/Subtractions        &Multiplications/Divisions       \\
 \hline
 OFDM symbol detection  &$6N_\mathrm{U}\sqrt{M}$    &$4N_\mathrm{U}\sqrt{M}$            \\ 
 \hline
 FFT/IFFT               &  $5\left((N/2)\log_2{N}\right) + 2N\log_2{N}$                                                                                                          &$3\left((N/2)\log_2{N}\right)$\\
 \hline
 Equalization           &$5N_\mathrm{U}$                &$3N_\mathrm{U}$          \\
 \hline
 SISO Precoding         &$5N_\mathrm{U}$                &$3N_\mathrm{U}$             \\
 \hline
 SISO Propagation       &$5N_\mathrm{U}$                &$3N_\mathrm{U}$           \\
 \hline
 SISO Nonlinearity      &$70N$                   &$5N$             \\ 
 \hline
\end{tabular}
\egroup
\end{table}
% \begin{table}[htb]
% \centering
% \caption{Number of operations of selected signal processing steps per single data subcarrier.}
% \label{table:num_operations_dsp}
% \bgroup
% \def\arraystretch{1.2}
% \begin{tabular}{ m{.23\columnwidth}  >{\centering\arraybackslash}  m{.31\columnwidth}  >{\centering\arraybackslash} m{.31\columnwidth}} 
%  \hline
%  \multirow{2}{*}{\begin{minipage}{.23\columnwidth}Signal processing step\end{minipage}} & \multicolumn{2}{c}{Operation count}  \\  \cline{2-3}
%                         & Additions/Subtractions        &Multiplications/Divisions       \\
%  \hline
%  OFDM symbol detection  &$6\sqrt{M}$    &$4\sqrt{M}$            \\ 
%  \hline
%  FFT/IFFT               &  $5\left((J/2)\log_2{\left(J N_{\mathrm{U}} \right)}\right) + 2J\log_2{\left(J N_{\mathrm{U}} \right)}$                 &$3\left((J/2)\log_2{\left(J N_{\mathrm{U}} \right)}\right)$\\
%  \hline
%  Equalization           &$5$                &$3$          \\
%  \hline
%  SISO Precoding         &$5$                &$3$             \\
%  \hline
%  SISO Propagation       &$5$                &$3$           \\
%  \hline
%  SISO Nonlinearity      &$70J$                   &$5J$             \\ 
%  \hline
% \end{tabular}
% \egroup
% \end{table}
The computational complexity of considered receivers is shown in Tab. \ref{tab:rx_complexity}.
\begin{table*}[ht]
  \centering
  \caption{Computational complexity of considered receivers.}
  \label{tab:rx_complexity}
  \bgroup
  \def\arraystretch{1.2}
  \begin{tabular}{>{\centering\arraybackslash} m{.2\textwidth}>{\centering\arraybackslash} m{.3\textwidth}>{\centering\arraybackslash} m{.35\textwidth}}
    \hline
    Receiver & Additions/Subtractions & Multiplications/Divisions \\ 
    \hline
    Standard OFDM receiver: equalization, FFT and detection $(0^{\text{th}} \ \text{ITE})$ 
            & $5N_\mathrm{U} + 5((N/2)\log_2{N})+2N\log_2{N} + 6N_\mathrm{U}\sqrt{M}$ 
            & $3N_\mathrm{U} + 3((N/2)\log_2{N}) + 4N_\mathrm{U}\sqrt{M}$ \\
    \hline
    Clipping Noise Cancellation (CNC) receiver 
            & $(0^{\text{th}} \ \text{ITE A/S}) + I\Big( 2\big(5((N/2)\log_2{N})+ 2N\log_2{N}\big) + 70N + 2N_\mathrm{U} + 6N_\mathrm{U}\sqrt{M} \Big)$ 
            & $(0^{\text{th}} \ \text{ITE M/D}) + I\Big( 2\big(3((N/2)\log_2{N})\big) + 5N + 2N_\mathrm{U} + 4N_\mathrm{U}\sqrt{M} \Big)$ \\
    \hline
    Multi-antenna Clipping Noise Cancellation (MCNC) receiver 
            & $ (0^{\text{th}} \ \text{ITE A/S}) + I\Big( (K+1)\big(5((N/2)\log_2{N})  
            + 2N\log_2{N}\big) + 70KN + (2K+1)(5N_\mathrm{U}) + (K-1)N_\mathrm{U} + 2N_\mathrm{U} + 6N_\mathrm{U}\sqrt{M} \Big)$ 
            & $ (0^{\text{th}} \ \text{ITE M/D}) + I\Big( (K+1)\big(3((N/2)\log_2{N})\big)  + 5KN + (2K+1)3N_\mathrm{U} + 4N_\mathrm{U}\sqrt{M}\Big)$ \\
    \hline
  \end{tabular}
\egroup
\end{table*}
Table \ref{tab:total_num_of_operations_calc} presents the total number of arithmetic operations required for a given number of iterations of the CNC and MCNC algorithm for $M = 64, N = 4096, N_\mathrm{U} = 2048, K = 64$. The values presented for the 0-th iteration correspond to the standard receiver, which performs equalization and demodulation. It can be seen that the complexity of the MCNC algorithm grows rapidly with the number of iterations and is substantially higher due to individual signal processing for each of the transmit antennas in the system. On the other hand, CNC algorithm complexity is relatively close to the standard receiver, which may advocate its application. 
Keep in mind that the additional arithmetical operations, in relation to the standard OFDM receiver, will cause OFDM symbol reception delay dependent on the computational capabilities of the receiver. 

%Due to the computational complexity of the algorithms, their execution time will be longer than the standard receiver. This consequent delay will depend upon the total number of arithmetic operations and the computational power of the receiver. Given the parameters and computational capability of the receiver, the delay in regard to the standard reception can be estimated based on Tab. \ref{tab:rx_complexity}.

\begin{table}[htb]
\centering
\caption{Total number of operations for $M = 64, N = 4096, N_{\mathrm{U}}=2048, K = 64$ and a selected number of iterations of the CNC and MCNC algorithms.}
\label{tab:total_num_of_operations_calc}
\bgroup
\def\arraystretch{1.2}
\begin{tabular}{ >{\centering\arraybackslash} m{.14\columnwidth}  >{\centering\arraybackslash}  m{.14\columnwidth}  >{\centering\arraybackslash} m{.14\columnwidth} >{\centering\arraybackslash} m{.14\columnwidth} >{\centering\arraybackslash} m{.14\columnwidth}}
 \hline 
  & \multicolumn{4}{c}{Total number of operations per data subcarrier $(10^3)$ } \\ \cline{2-5}
 \multirow{2}{*}{\begin{minipage}{.17\columnwidth}Number of \\ iterations: $I$\end{minipage}} & \multicolumn{2}{c}{Additions/subtractions} & \multicolumn{2}{c}{Multiplications/divisions} \\ \cline{2-5}
    & CNC & MCNC & CNC & MCNC \\ \cline{1-5}
    0 & 0.16 & 0.16 & 0.07 & 0.07 \\ 
    \hline
    1 & 0.57 & 16.84 & 0.19 & 3.47 \\ 
  %  \hline
   % 2 & 0.97 & 33.51 & 0.30 & 6.87 \\ 
    \hline
    3 & 1.38 & 50.19 & 0.42 & 10.27 \\ 
    \hline
%    4 & 1.78 & 66.86 & 0.54 & 13.67 \\ 
 %   \hline
  %  5 & 2.19 & 83.54 & 0.65 & 17.07 \\ 
   % \hline
%    6 & 2.60 & 100.21 & 0.77 & 20.46 \\ 
 %   \hline
 %   7 & 3.00 & 116.89 & 0.88 & 23.86 \\ 
  %  \hline
    8 & 3.41 & 133.56 & 1.00 & 27.26 \\ 
    \hline
\end{tabular}
\egroup
\end{table}

\section{Simulation results}
\label{sec:simulation_results}
The performance of considered clipping noise cancellation algorithms is evaluated by computer simulations. The transmitting end is a uniform linear array with an inter-element spacing of half wavelength. Each antenna is modeled as an omnidirectional radiator with a gain of 0 dBi. The transmitter end was positioned 15 m above the ground level. Tab. \ref{tab:sim_params} presents the details concerning the simulation setup. Each front-end amplifier was modeled as a soft limiter with identical cutoff power. The receiver was placed 300 m from the TX at an azimuth of 45\degree \ and 1.5m above the ground level. If not stated differently, perfect channel state information is available both at the transmitter and receiver. The transmitter employs MRT precoding. We consider mostly 3 types of radio channels: 1)
%\begin{itemize}
%    \item 
    LOS: modeled as an attenuation of the free space and phase rotation resulting from the distance between each transmitting antenna and the receiver;
    %\item 
    2) Two-path: apart from the direct path it includes an additional one corresponding to the reflection from the ground with a reflection coefficient equal to $-1$. The point of reflection is calculated taking into consideration the location of the receive (RX) and TX elements;
    %\item 
    3) Rayleigh: modeled as independent, identically distributed complex Gaussian variables for each subcarrier and antenna.
%\end{itemize}
Each result is obtained after transmitting approximately 800 OFDM symbols with independent modulating symbols. For the Rayleigh channel, each symbol is transmitted through an independently generated channel. For the LOS and two-path channels for each symbol, the position of the receiver is picked randomly within a 10m square centered at the reference position.

\begin{table}[htb]
\centering
\caption{Simulation parameters}
\label{tab:sim_params}
\begin{tabular}{llll}
\hline
\textit{Parameter} & \textit{Symbol} & \textit{Value} & \textit{Unit} \\ \hline 
Subcarrier spacing                   & $\Delta f$        & $15$      & $\left[\mathrm{kHz}\right]$   \\
Carrier frequency                    & $f_c$             & $3.5$     & $\left[\mathrm{GHz}\right]$   \\
Total number of subcarriers          & $N$               & $4096$    & $\left[-\right]$              \\
Number of data subcarriers           & $N_\mathrm{U}$    & $2048$    & $\left[-\right]$              \\
QAM constellation size               & $M$               & $64$      & $\left[-\right]$              \\
Number of iterations CNC/MCNC        & $I$               & $0-8$      & $\left[-\right]$              \\
\hline
\end{tabular}
\end{table}

\subsection{Results}
\begin{figure}[htb]
\centering
\includegraphics[width=3.6in]{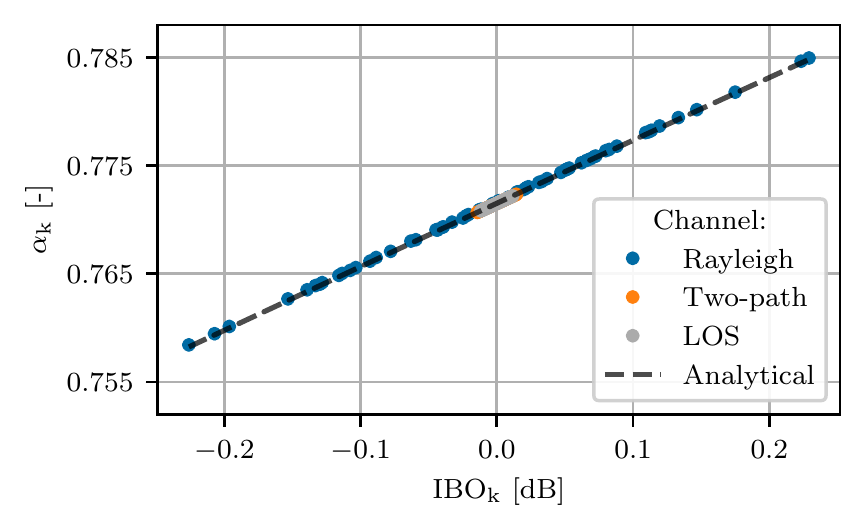}
\caption{$\mathrm{IBO}_k$ and $\alpha_k$ values of individual antenna front-ends for $K = 64$ and selected channels.}
\label{fig:alpha_ibo_per_ant}
\end{figure}

First, we show in Fig. \ref{fig:alpha_ibo_per_ant} values of estimated and analytical $\alpha_k$ with respect to $\mathrm{IBO}_k$ for $\mathrm{IBO}=0$ dB. Recall that $\mathrm{IBO}_k$ is IBO calculated individually for each TX antenna considering the utilized precoding vectors. It is visible that for all considered channels the $\alpha_k$ values vary slightly among front-ends. Most importantly, in all the cases the estimated $\alpha_k$ value follows the analytical result of (\ref{eq:analytical_alpha}) as discussed in Sec. \ref{sec_Nonlinear_amplifier}. The value of $\alpha_k$ depends only on $\mathrm{IBO}$ of each individual front-end. 

\begin{figure}[htb]
\centering
\includegraphics[width=3.6in]{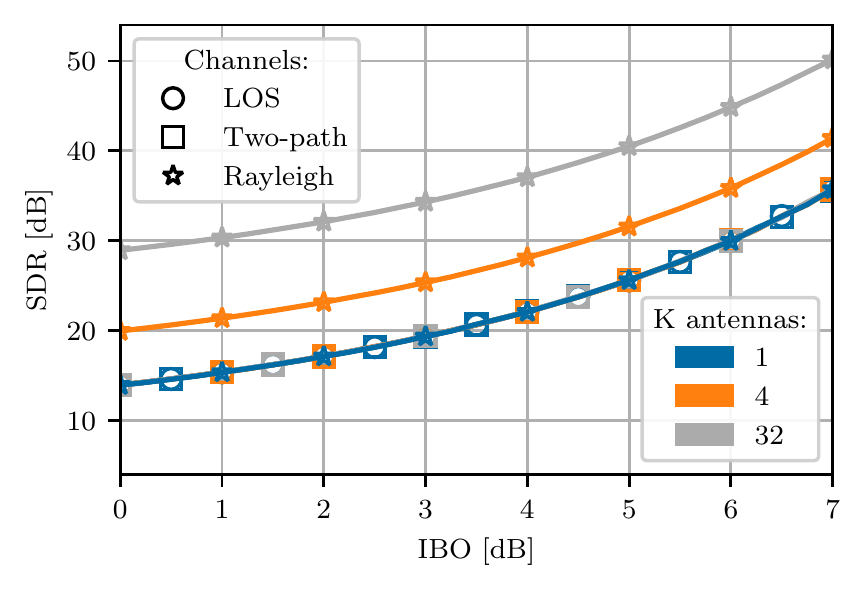}
\caption{SDR with respect to IBO for selected channels and a number of antennas.}
\label{fig:sdr_vs_ibo}
\end{figure}

Next, the signal-to-distortion ratio was plotted against the IBO for selected channels as shown in Fig. \ref{fig:sdr_vs_ibo}. While the MRT precoding is expected to provide $10\log_{10}(K)$ dB gain of the wanted signal, at the same time it can increase the power of nonlinear distortion arriving at the receiving antenna \cite{oob_clarified_two_tone_test}. This happens both for LOS and two-path channels as increasing the number of antennas does not change the SDR value. Only for the considered Rayleigh channel, the nonlinear distortion can be reduced by increasing $K$ as expected in \cite{bjornson_nonlinearity_as_white_noise}. However, keep in mind that the considered Rayleigh channel model is independent and identically distributed both among antennas and subcarriers. A similar effect can be observed if multiple users are served in parallel, i.e., this improves the SDR performance with respect to single-user precoding \cite{mollen_spatial_char}. This shows that while utilization of a massive number of antennas can combat many phenomena, e.g., high path-loss or channel fadings, there is still in some scenarios a need for solutions removing the impact of nonlinear PAs. We consider single-user precoding as the most challenging from a nonlinear distortion perspective.

\begin{figure}[htb]
\centering
\includegraphics[width=3.6in]{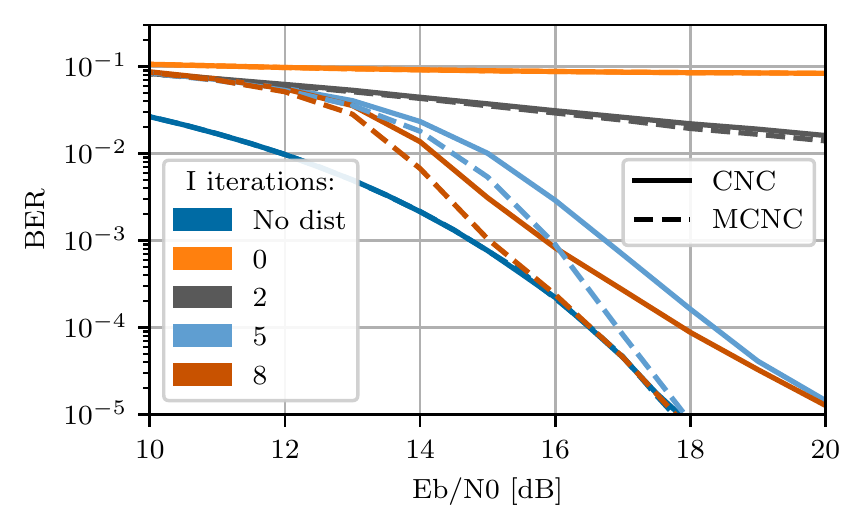}
\caption{BER vs Eb/N0 for IBO = 0 dB, $K=64$ antennas, LOS channel and a selected number of iterations of the CNC and MCNC algorithm.}
\label{fig:ber_vs_ebn0_los}
\end{figure}

\begin{figure}[htb]
\centering
\includegraphics[width=3.6in]{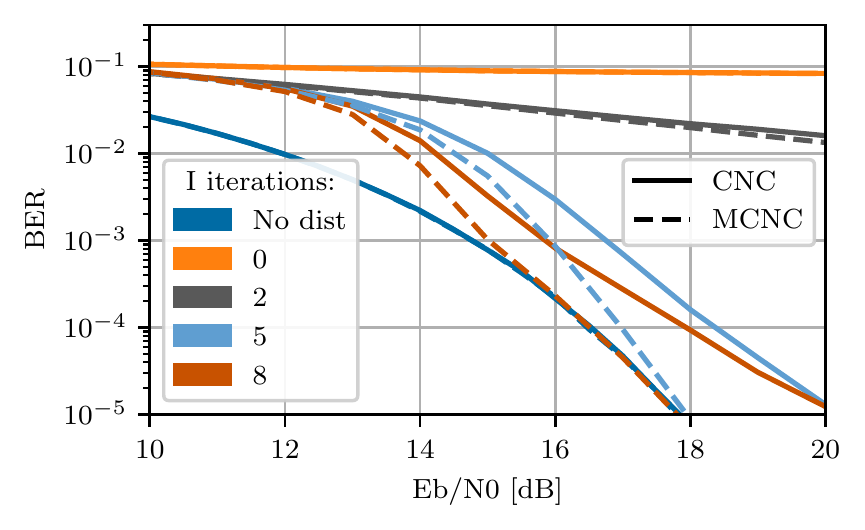}
\caption{BER vs Eb/N0 for IBO = 0 dB, $K=64$ antennas, two-path channel and a selected number of iterations of the CNC and MCNC algorithm.}
\label{fig:ber_vs_ebn0_two_path}
\end{figure}

\begin{figure}[htb]
\centering
\includegraphics[width=3.6in]{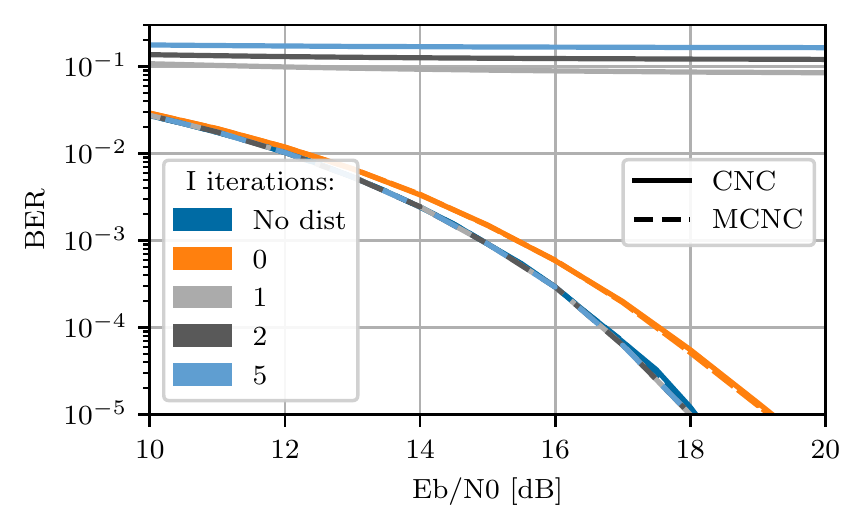}
\caption{BER vs Eb/N0 for IBO = 0 dB, $K=64$ antennas, Rayleigh channel and a selected number of iterations of the CNC and MCNC algorithm.}
\label{fig:ber_vs_ebn0_rayleigh}
\end{figure}

\begin{figure}[htb]
\centering
\includegraphics[width=3.6in]{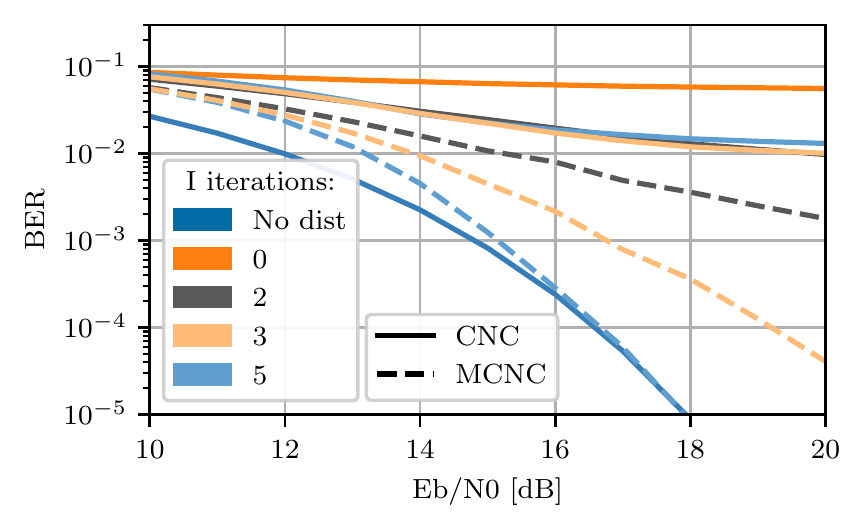}
\caption{BER vs Eb/N0 for IBO = 0 dB, $K=64$ antennas, 38.901 Urban Macrocell LOS channel and a selected number of iterations of the CNC and MCNC algorithm.}
\label{fig:ber_vs_ebn0_uma_los}
\end{figure}

\begin{figure}[htb]
\centering
\includegraphics[width=3.6in]{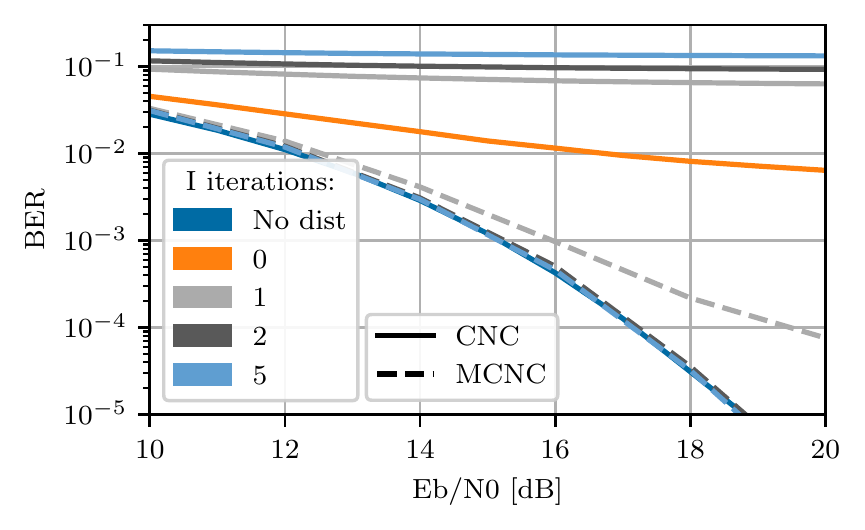}
\caption{BER vs Eb/N0 for IBO = 0 dB, $K=64$ antennas, 38.901 Urban Macrocell NLOS channel and a selected number of iterations of the CNC and MCNC algorithm.}
\label{fig:ber_vs_ebn0_uma_nlos}
\end{figure}

In order to present gains from MCNC and CNC methods, we start by fixing IBO to 0 dB (significant nonlinear distortion), $K$ to 64,  and testing BER for varying Eb/N0 and a number of RX iterations. The results for LOS, two-path,  Rayleigh, and 3GPP 38.901 Urban Macrocell LOS, and NLOS \cite{3gpp_nr_channel_model} channels are presented in Fig. \ref{fig:ber_vs_ebn0_los}, \ref{fig:ber_vs_ebn0_two_path}, \ref{fig:ber_vs_ebn0_rayleigh}, \ref{fig:ber_vs_ebn0_uma_los}, and \ref{fig:ber_vs_ebn0_uma_nlos}, respectively. The 3GPP channels are generated using Quadriga \cite{quadriga_paper}. First, it is visible that results for LOS and two-path channels are very close to each other in all considered scenarios, revealing significant distortions level resulting in BER close to $10^{-1}$ for standard RX in the whole observation range. This shows, similarly to Fig. \ref{fig:sdr_vs_ibo}, that not only a LOS channel, as shown in \cite{oob_clarified_two_tone_test}, but also a sparse multi-path channel can suffer from nonlinear distortion in mMIMO systems. Observe that in the case of the Rayleigh channel the directly received distorted signal (0th iteration) achieves much lower BER for the same Eb/N0 in relation to LOS or a two-path channel. This is the result of antenna array gain improving SDR as has been shown in Fig. \ref{fig:sdr_vs_ibo}. Secondly, for all considered channels MCNC allows to achieve the BER limit observed for a system without nonlinear distortion (\emph{No dist} in figures) for high Eb/N0 after no more than 8 iterations. The BER improvement increases with the number of RX iterations. However, this happens at the cost of significant computational complexity as the receiver has to emulate the signal processing of all considered TX-RX links. 
% The CNC algorithm requires 8 iterations to achieve this BER equal to 16 dB. Meanwhile, the MCNC algorithm achieves the same BER given the same Eb/N0 using only 5 iterations. However, the cost of 8 iterations of CNC is 9.64 million additions/subtractions and 3.82 million multiplications/divisions compared to 172.85 and 36.13 million of respective operations for 5 iterations of MCNC.
Significantly lower computational complexity and a lower amount of control information are required by the CNC algorithm. As visible in Fig. \ref{fig:ber_vs_ebn0_los}, and Fig. \ref{fig:ber_vs_ebn0_two_path} the CNC algorithm allows for significantly improved BER for LOS and two-path channels. However, the performance is slightly worse than for the MCNC algorithm. After the 8th iteration for BER = $10^{-5}$ the loss equals about 2 dB in Eb/N0. 
%Though, for a fair comparison, let us consider fixed BER of $10^{-3}$ for both algorithms next. 
For the considered Rayleigh channel the utilization of the MCNC algorithm results in \emph{No dist} performance. On the other hand the CNC algorithm increases BER. While there is an independent random channel coefficient on each subcarrier for each TX antenna, the MRT precoding coefficient varies similarly influencing samples of nonlinear distortion, i.e., $\sum^K_{k=1} h_{k,n}d_{k,n}$ in (\ref{eq:rx_sig}). While the CNC algorithm is unaware of the precoding it is reconstructing the clipping noise that is significantly different than the real one deteriorating reception performance.  
%The performance of the CNC and MCNC algorithm was also evaluated for practical channels specified in \cite{3gpp_nr_channel_model} with the use of Quadriga channel generator \cite{quadriga_paper}. For the evaluation, two channel scenarios were selected Urban Macrocell LOS and NLOS. 
Fig. \ref{fig:ber_vs_ebn0_uma_los} shows the BER vs Eb/N0 curve with the 3GPP Urban Macrocell LOS channel. It can be seen that the CNC algorithm still offers improvement in regard to the standard RX, though, due to frequency selective fading its gains are significantly limited. The MCNC takes into consideration the fading and is able to efficiently remove the distortion with a few iterations obtaining \emph{No dist} performance for higher Eb/N0 values.
The results for the NLOS version of the 3GPP channel are shown in Fig. \ref{fig:ber_vs_ebn0_uma_nlos}. The NLOS case can be observed to exhibit some SDR increase by the array gain as the $0$-th iteration curve is lower than in the 3GPP LOS case. Similarly to the ideal Rayleigh channel the CNC algorithm does not work and MCNC needs only a few iterations to reach the floor corresponding to the no distortion case. 
%In this case, the channels between antennas are less correlated than in the LOS case and the CNC algorithm fails to cancel the distortion. Similarly to the Rayleigh channel, only the MCNC algorithm is able to improve the performance and after 2 iterations it matches the performance without nonlinear distortion, due to SDR gains caused by partially uncorrelated channels.

\begin{figure}[htb]
\centering
\includegraphics[width=3.6in]{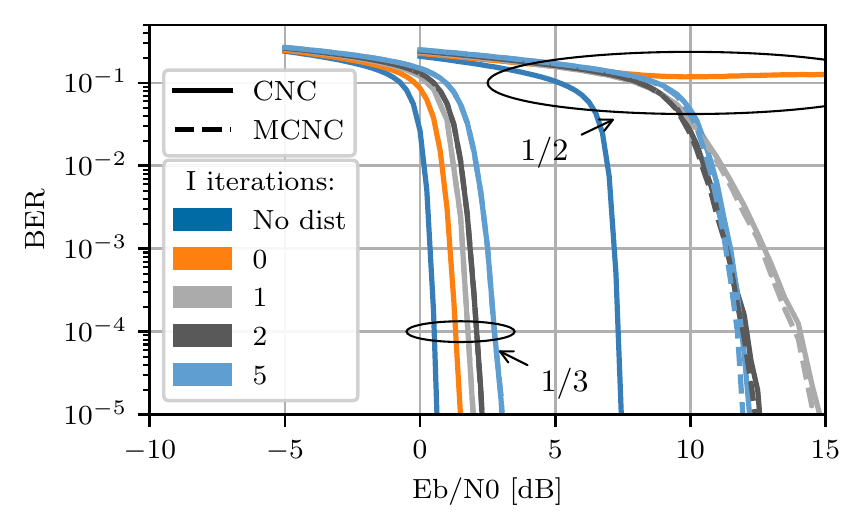}
\caption{BER vs Eb/N0 for IBO = 0 dB, K = 64 antennas, LOS channel, two code rates of LDPC channel coding and a selected number of iterations of the CNC and MCNC algorithm.}
\label{fig:ldpc_ber_vs_ebn0_los}
\end{figure}

%Typically OFDM schemes are combined with channel coding, with substantially different coded and uncoded performances. 
Next, the CNC and MCNC algorithms were evaluated in the presence of 5G NR-compliant low-density parity check (LDPC) coding \cite{3gpp_5g_nr_channel_coding}. The coding and decoding is performed with the use of Matlab nrDLSCH package \cite{matlab_nrdlsch_package}. Utilized LDPC coding follows 5G NR Shared Channel processing, e.g., embedding cyclic redundancy check (CRC) bits. The code parameters before the rate matching are as follows: single code block, 104 filler bits, 192 lifting size, 4224 bits per code block, and 12672 bits per code block after LDPC coding for code rate 1/3, and single code block, 232 filler bits, 384 lifting size, 8448 bits per code block, and 25344 bits per code block after LDPC coding for code rate 2/3. The decoding algorithm is the belief propagation. Figure \ref{fig:ldpc_ber_vs_ebn0_los} shows the BER curves of the CNC and MCNC algorithms for two code rates of 1/2 and 1/3 in the LOS channel. The algorithms do not offer any gains for the lower code rate (1/3) and each iteration increases the error rate. This is caused by the LDPC decoder having a waterfall region before the CNC/MCNC algorithms start to improve signal quality on the LDPC decoder input. For the higher code rate, both CNC and MCNC algorithms provide significant quality improvement with respect to the standard RX (0th iteration). As such the proposed CNC/MCNC algorithms can be useful for a coded system but require wise modulation and coding scheme selection for a given nonlinearity and channel distortion conditions. The scheme might be further improved by introducing the LDPC decoder and encoder inside the MCNC/CNC loop as in \cite{ldpc_inside_cnc_loop_optical_coms}. 

\begin{figure}[htb]
\centering
\includegraphics[width=3.6in]{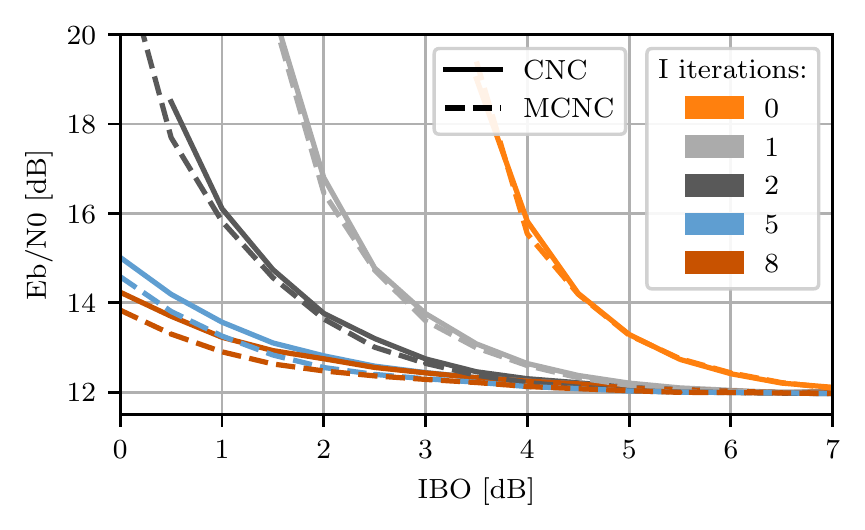}
\caption{Eb/N0 vs IBO for a fixed BER = $10^{-2}$, $K=64$ antennas, two-path channel and a selected number of iterations of the CNC and MCNC.}
\label{fig:fixed_ber_vs_ibo_vs_ebn0_two_path}
\end{figure}

\begin{figure}[htb]
\centering
\includegraphics[width=3.6in]{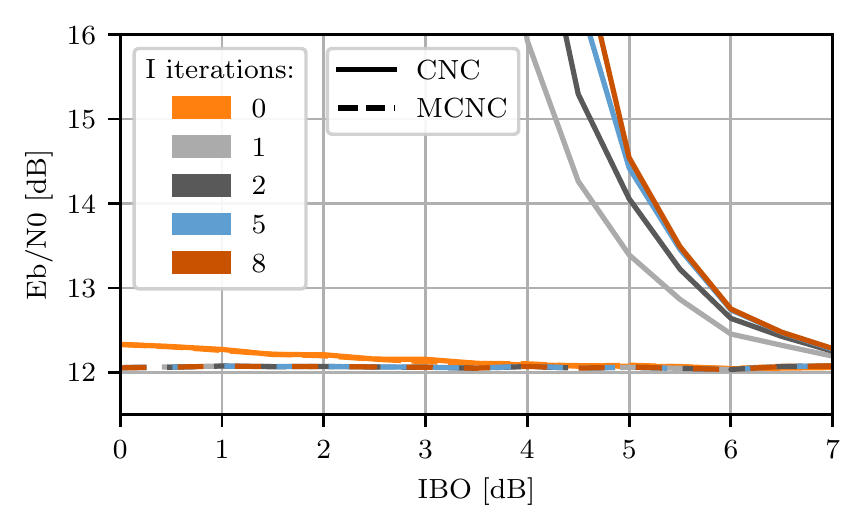}
\caption{Eb/N0 vs IBO for a fixed BER = $10^{-2}$, $K=64$ antennas, Rayleigh channel and a selected number of iterations of the CNC and MCNC.}
\label{fig:fixed_ber_vs_ibo_vs_ebn0_rayleigh}
\end{figure}

Next, the proposed RX algorithms are tested for varying PA operating points, i.e., IBO.
%The Eb/N0 is fixed to 15 dB and the number of antennas is $K=64$. BER curves for two-path and Rayleigh channels are shown in Fig. \ref{fig:ber_vs_ibo_two_path}, and Fig. \ref{fig:ber_vs_ibo_rayleigh}, respectively. The results for the LOS channel are omitted as are identical to the two-path channel. The results confirm the previous observations. The MCNC algorithm achieves a noise-limited BER floor around $10^{-3}$ after a sufficiently high number of iterations. The lower the clipping level the more iterations are needed. The CNC algorithm provides BER improvement with respect to the standard receiver (0th iteration) only for two-ray and LOS channels being slightly outperformed by the MCNC algorithm.  
%\begin{figure}[htb]
%\centering
%\includegraphics[width=3.6in]{plots/ber_vs_ibo_two_path_nant64_ebn0_15_ibo_min0_max8_step0.50_niter1_2_5_8.pdf}
%\caption{BER vs IBO for Eb/N0 = 15 dB, $K=64$ antennas, Two-path channel and a selected number of iterations of the CNC and MCNC.}
%\label{fig:ber_vs_ibo_two_path}
%\end{figure}
%\begin{figure}[htb]
%\centering
%\includegraphics[width=3.6in]{plots/ber_vs_ibo_rayleigh_nant64_ebn0_15_ibo_min0_max8_step0.50_niter1_2_5_8.pdf}
%\caption{BER vs IBO for Eb/N0 = 15 dB, $K=64$ antennas, Rayleigh channel and a selected number of iterations of the CNC and MCNC.}
%\label{fig:ber_vs_ibo_rayleigh}
%\end{figure}
Figure \ref{fig:fixed_ber_vs_ibo_vs_ebn0_two_path} and \ref{fig:fixed_ber_vs_ibo_vs_ebn0_rayleigh} visualize the gains of the CNC and MCNC algorithm for a fixed BER value equal to $10^{-2}$ in regard to both Eb/N0 and IBO. This form of presentation allows to evaluate the gains from using a specific number of iterations. Given the IBO it is possible to estimate the margin by which the Eb/N0 requirements can be reduced for a certain number of iterations and vice versa. For direct visibility channels: LOS and two-path only the results for two-path are shown as the results are highly identical and differ only up to the accuracy of the simulations. In Fig. \ref{fig:fixed_ber_vs_ibo_vs_ebn0_two_path} it can be observed that for these channels the gains from using the MCNC algorithm over standard CNC become apparent since the second iteration. The lower the IBO the higher number of iterations required to meet the Eb/N0 12 dB floor which corresponds to the system without nonlinear distortion. For the Rayleigh channel and MCNC reception in Fig. \ref{fig:fixed_ber_vs_ibo_vs_ebn0_rayleigh} required Eb/N0 curve is almost flat for any value of IBO from the range. The first iteration of the MCNC offers minimal improvement. This is due to a high number of antennas $K=64$, which translates into higher SDR in the Rayleigh channel, as could be seen in Fig. \ref{fig:sdr_vs_ibo}, lessening the severity of the impact of nonlinear distortion on the received signal and allowing the algorithm to work with less nonlinear distortion interference.
% \begin{figure}[htb]
% \centering
% \includegraphics[width=3.6in]{plots/fixed_ber1.0e-02_los_nant64_ebn0_min10_max22_step0.50_ibo_min0_max7_step0.50_niter1_2_5_8.pdf}
% \caption{Eb/N0 in regard to IBO for a fixed BER = $10^{-2}$, 64 antennas, LOS channel and selected number of iterations of the CNC and MCNC.}
% \label{fig:fixed_ber_vs_ibo_vs_ebn0_los}
% \end{figure}

\begin{figure}[htb]
\centering
\includegraphics[width=3.6in]{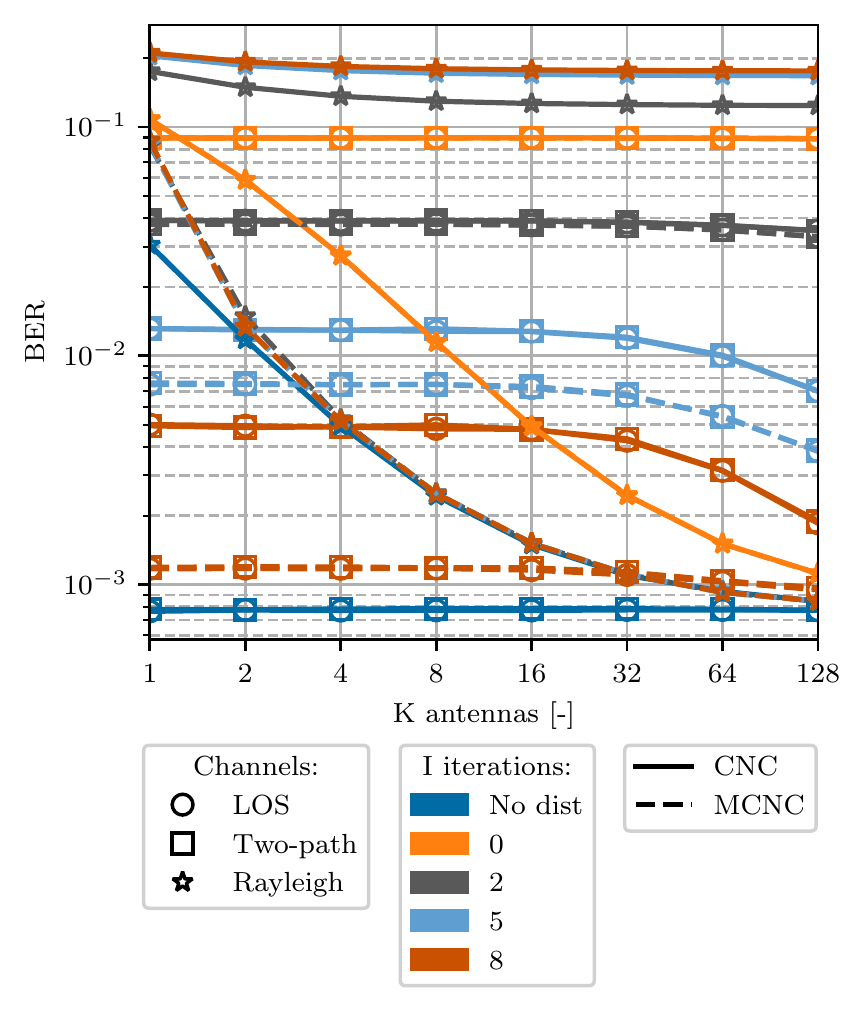}
\caption{BER vs the number of antennas $K$, for LOS, two-path, and Rayleigh channels, for Eb/N0 = 15 dB, IBO = 0 dB and a selected number of iterations of the CNC, and MCNC.}
\label{fig:ber_vs_nant_vs_chan}
\end{figure}
Figure \ref{fig:ber_vs_nant_vs_chan} presents a comparison between CNC and MCNC algorithms taking into consideration the channel type, number of RX iterations, and number of antennas $K$. The first observation can be a significant decrease in BER for the Rayleigh channel with the number of antennas. This effect is due to precoding gains which increase the SDR with the number of antennas as $10\log_{10}{(K)}$. As expected from previous results, while the MCNC helps to improve the BER performance, the CNC algorithm increases BER in this scenario. For a high number of antennas in the Rayleigh channel, the SDR gains allow the MCNC algorithm to quickly converge within a single iteration to the noise-limited bound denoted as \emph{No dist}. On the other hand, the CNC algorithm works well for LOS and two-path channels achieving BER slightly higher than the MCNC algorithm. Again, the performance of LOS and two-path channels is nearly identical. An interesting observation for these channels is that while the BER performance for both iterative RX algorithms remains constant up to about $K=16$ antennas it starts to slightly decrease for greater $K$ and a greater number of RX iterations. For a high number of iterations, e.g, 8, this phenomenon vanishes, with the MCNC algorithm performing close to the noise-limited bound.

\begin{figure}[htb]
\centering
\includegraphics[width=3.6in]{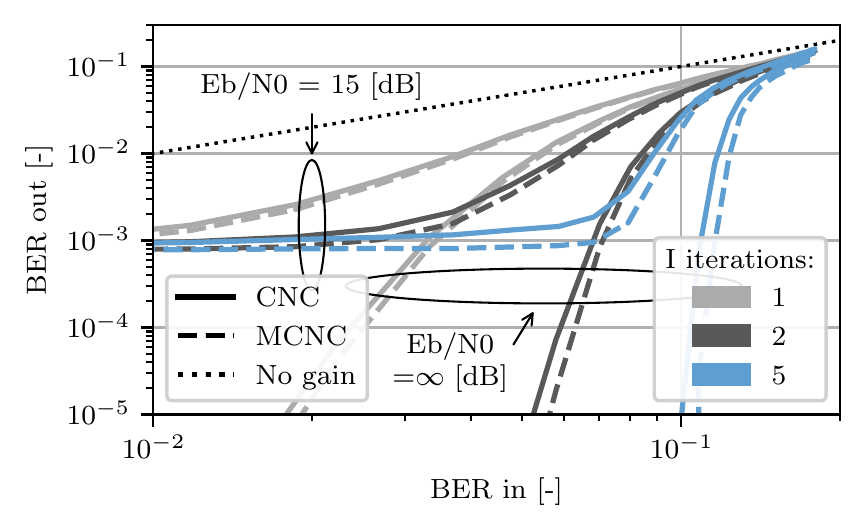}
\caption{BER out vs BER in after $I$ iterations, $K=64$ antennas, LOS channel, varying IBO for selected values of Eb/N0 and MCNC/CNC iterations.}
\label{fig:berin_vs_berout_per_combined_ite}
\end{figure}

% The convergence of each iteration of the CNC and MCNC algorithm is presented in Fig. \ref{fig:berin_vs_berout_per_ite}. The different BER input values were obtained by sweeping the IBO value, the higher BER at input corresponds to lower IBO and greater nonlinear distortion. The figure shows that for high BER in, which translates to low nonlinear distortion the performance of the CNC and MCNC receiver is ultimately limited by noise. The BER floor corresponding to the no distortion case of Eb/N0 = 15 dB is marked by a blue dot. For severe nonlinear distortion, which corresponds to the low BER in values and no noise present (Eb/N0 = $\infty$) the performance of the algorithms is also constrained due to irrecoverable symbol distortion. The BER input value for which the curves start to deviate from the no-gain diagonal can be considered the BER threshold from which the algorithms start to converge.
Figure \ref{fig:berin_vs_berout_per_combined_ite} presents BER after $I$ iterations of CNC and MCNC algorithm (BER out) as a function of BER on the input, i.e., obtained with a standard receiver (BER in). Two values of Eb/N0 are tested while varying IBO values resulting in a range of input BER values. The closer a given result of the CNC/MCNC algorithm is to the \emph{no gain} line the smaller BER improvement is obtained. It is visible that in the case of Eb/N0 of 15 dB the system cannot reduce output BER below around $10^{-3}$, being the noise-caused error level. As expected, increasing the number of iterations reduces in most cases the achievable output BER. This effect is more significant when the nonlinear distortion is the dominating distortion in the system, e.g., here for Eb/N0 equal $\infty$. 
Most importantly, the BER in value for which the curves start to deviate from the no-gain diagonal can be considered as a BER threshold from which the CNC/MCNC algorithms start to \emph{work}. In this case it is around BER in of $10^{-1}$. 
% \begin{figure}[htb]
% \centering
% \includegraphics[width=3.6in]{plots/berout_vs_berin_per_single_ite_los_nant64_ebn015_1000_ibo_min-9_max9_step0.50_niter1_2_5.pdf}
% \caption{BER in vs BER out per single $I$-th iteration, $K=64$ antennas, LOS channel, varying IBO for selected values of Eb/N0 and MCNC and CNC iterations.}
% \label{fig:berin_vs_berout_per_ite}
% \end{figure}

\begin{figure}[htb]
    \centering
    \includegraphics[width=3.5in]{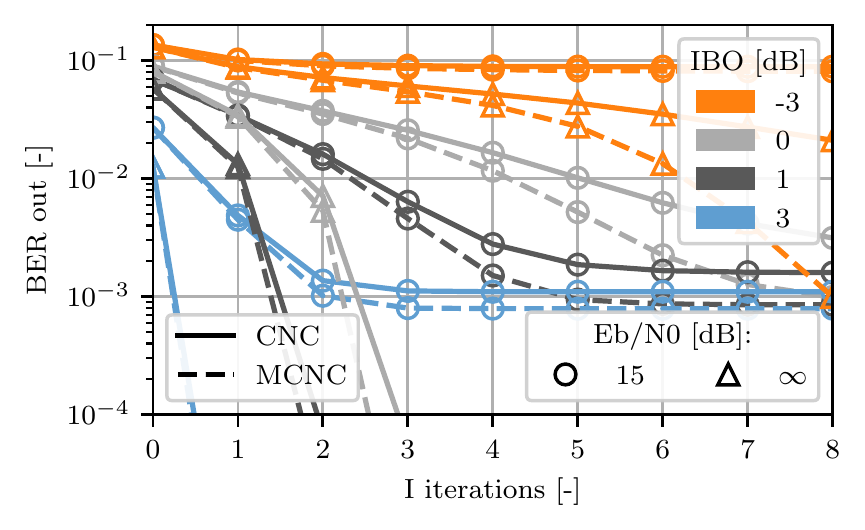}
    \caption{BER out vs the number of $I$ iterations of MCNC or CNC algorithm, $K=64$ antennas, LOS channel and selected values of IBO and Eb/N0.}
    \label{fig:berout_per_ite}
\end{figure}

Figure \ref{fig:berout_per_ite} presents the evolution of BER at the output of the CNC/MCNC algorithms as a function of a number of iterations. It is visible that for a given Eb/N0 value the CNC/MCNC algorithms converge the faster the lower nonlinear distortion power is present. The convergence is slightly faster for the MCNC algorithm. Moreover, the lower the thermal noise the faster convergence is possible.

\begin{figure}[htb]
    \centering
    \includegraphics[width=3.5in]{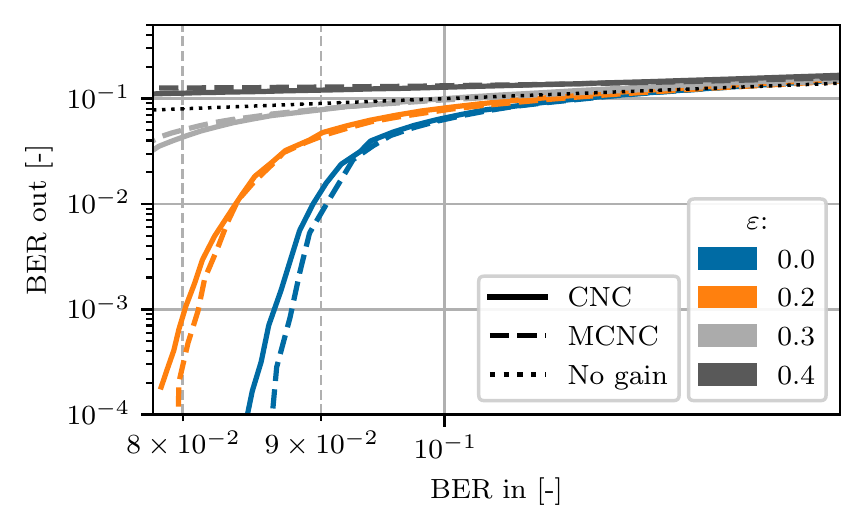}
    \caption{BER after $I=5$ iterations of MCNC or CNC algorithm as a function of input BER (for a standard receiver), $K=64$ antennas, IBO = 0 dB and selected values of CSI estimation error.}
    \label{fig:ber_out_vs_berin_csi}
\end{figure}

Figure \ref{fig:ber_out_vs_berin_csi} presents the impact of the channel state information (CSI) error on the performance of the CNC and MCNC algorithms in an ideal LOS channel. The CSI error is modeled as in \cite{csi_err_model} with parameter $\varepsilon \in \langle 0; 1\rangle$ giving the estimated channel coefficient $\hat{h}_{k,n} = \sqrt{1-\varepsilon^2} h_{k,n} + \varepsilon w_{k,n}$,
where $w_{k,n}$ is the white noise sample with the power corresponding to the average gain of the channel for the data subcarriers $w_{k,n} = \mathcal{CN}(0,1) \sqrt{\frac{\sum_{n\in \mathcal{N}}{\left| h_{k,n} \right|}^2} {N_{\mathrm{U}}}}$
and $\mathcal{CN}(0,1)$ represents a complex normal variable with expected value 0 and variance 1. The inaccurate channel estimate denoted as $\hat{h}_{k,n}$ is used both at the base station for precoding and at the receiver within the MCNC algorithm loop.
With the increasing value of $\varepsilon$ the gains of the algorithms are smaller and shifted towards smaller values of BER in. The CNC and MCNC algorithms exhibit relatively high tolerance to channel estimation errors offering gains for $\varepsilon$ up to 0.3.

\begin{figure*}[htb]
        \centering
        \includegraphics[width=3.5in]{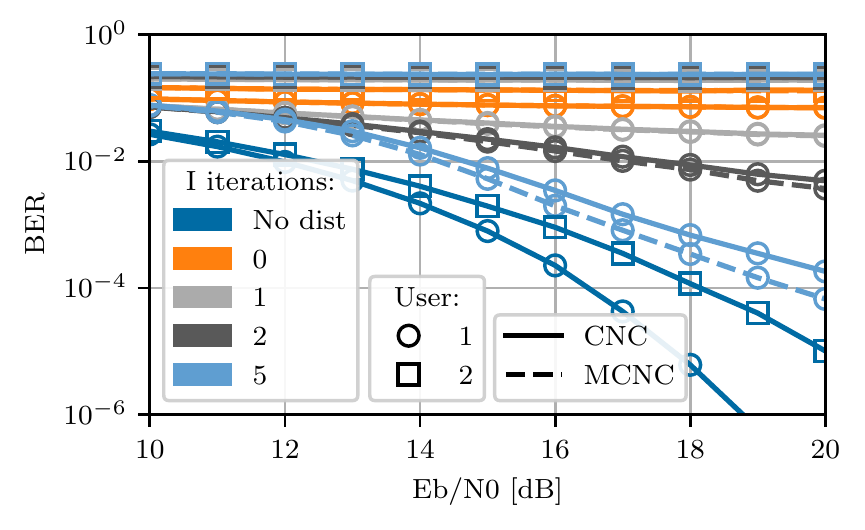}
        \caption{BER vs Eb/N0 for two users, $K=64$, IBO = 0 dB, LOS channel with MRT precoding.}
        \label{fig:nmu_cnc_mcnc_mr_preocding}
    \end{figure*}

Finally, the performance of the proposed CNC and MCNC receiver has been tested for a scenario with two users allocated at the same subcarriers. As explained in Sec. \ref{sec:multiuser}, the CNC/MCNC algorithms are still the single-user versions that treat the other user interference as noise. Fig. \ref{fig:nmu_cnc_mcnc_mr_preocding} presents the BER performance of the CNC and MCNC algorithms while using MRT precoding. The two users are located at azimuths -30\degree\ and 30\degree\ from the array. User 1 is located closer to the array and user 2 is further away with a path loss difference of 10 dB between them. MRT precoding allocates power to users proportionally to the channel magnitude. The reference, no-distortion curves differ between users due to different levels of inter-user interference. It is visible that BER reduction is obtained by CNC and MCNC only for user 1, while the CNC/MCNC algorithm increases BER for the other user. The failure of the CNC/MCNC algorithm comes from the inter-user interference, both its linear and nonlinear component, that the proposed algorithms do not remove. For user no. 1 the ratio between signal and interference power is higher, resulting in a lower BER value in iteration 0, enabling successful CNC/MCNC operation.       

\section{Conclusions}
\label{sec:conclusions}
It has been shown that the MRT precoding using a high number of antennas does not offer any SDR improvement in the presence of front-end nonlinearity for direct visibility channels, severely limiting the performance of the mMIMO system.
In this work, we have proposed the MCNC algorithm that is able to combat even severe nonlinear distortion in the downlink receiver of the mMIMO OFDM system. The system was tested for MRT precoding, single and two user scenarios and a few types of channels. While the MCNC algorithm is relatively complex and requires a high amount of information, its simplified version was introduced. The simulations have shown that for direct visibility channels: LOS and two-path the performance penalty of the simplified algorithm is not that substantial and it can be effectively utilized. 
An interesting future step would be to improve the mMIMO OFDM reception performance by leveraging the frequency diversity of nonlinear distortion as used for an OFDM system in \cite{belief_propagation_rx_ofdm}.

% \section*{Acknowledgment}
% This research was funded by the Polish National Science Centre, project no. 2021/41/B/ST7/00136. 

\bibliographystyle{IEEEtran}
\bibliography{biblio.bib}

\end{document}